\pdfoutput=1

\documentclass[11pt]{article}

\usepackage[final]{acl}

\usepackage{times}
\usepackage{latexsym}

\usepackage[T1]{fontenc}

\usepackage[utf8]{inputenc}

\usepackage{microtype}

\usepackage{inconsolata}

\usepackage{graphicx}

\usepackage{enumitem} 

\definecolor{red}{rgb}{1.00, 0.00, 0.00}

\definecolor{audio_color}{rgb}{0.8, 0.2, 0.3} 
\definecolor{human_color}{rgb}{0.3, 0.6, 0.5} 
\definecolor{machine_color}{rgb}{0.4, 0.6, 0.8}
\definecolor{cot_color}{RGB}{251,244,243}

\usepackage{hyperref}
\usepackage{tabularx, adjustbox, multirow, diagbox, makecell, booktabs}
\usepackage{amsmath}
\usepackage{soul}
\usepackage{bm}
\usepackage{tcolorbox}
\usepackage{amsfonts}
\usepackage{pifont}
\usepackage{array}

\usepackage{tikz}
\usepackage{pgfplots}
\pgfplotsset{compat=1.18}

\def\ear{{\includegraphics[width=0.02\textwidth]{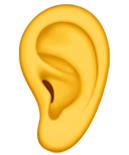}}{}}
\def\brain{{\includegraphics[width=0.02\textwidth]{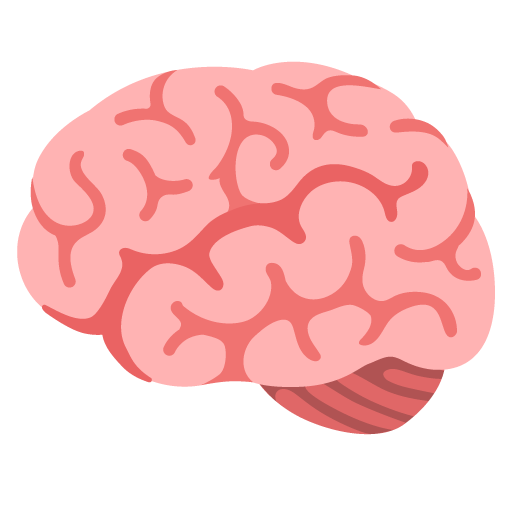}}{}}

\def\fire{{\includegraphics[width=0.025\textwidth]{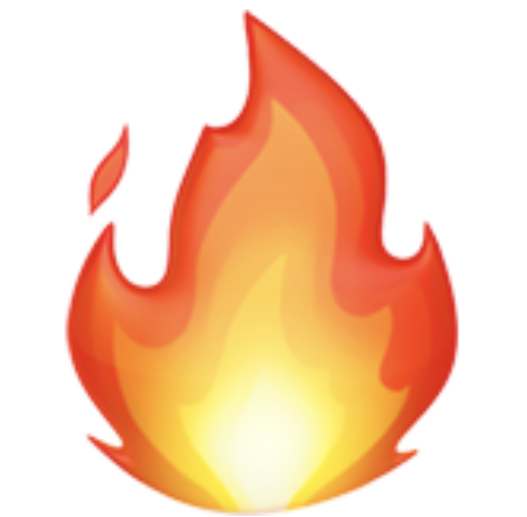}}{}}
\def\snow{{\includegraphics[width=0.025\textwidth]{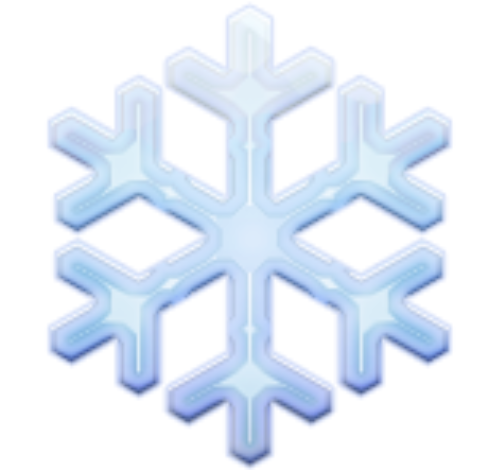}}{}}

\title{AAD-LLM: Neural Attention-Driven Auditory Scene Understanding}

\author{Xilin Jiang\textsuperscript{1, 3\ear}, Sukru Samet Dindar\textsuperscript{1, 3\brain}, Vishal Choudhari\textsuperscript{1, 3}, \\ {\bf Stephan Bickel\textsuperscript{4, 5}},{\bf Ashesh Mehta\textsuperscript{4, 5}}, {\bf Guy M McKhann\textsuperscript{2}}, \\ {\bf Daniel Friedman\textsuperscript{6}}, {\bf Adeen Flinker\textsuperscript{6}},  {\bf Nima Mesgarani\textsuperscript{1, 3}}\\
\{\textsuperscript{1}Department of Electrical Engineering, 
\textsuperscript{2}Department of Neurological Surgery, \\
\textsuperscript{3}Mortimer B. Zuckerman Mind Brain Behavior Institute\}, Columbia University, USA 
\\
\textsuperscript{4}Hofstra Northwell School of Medicine, USA \\
\textsuperscript{5}The Feinstein Institutes for Medical Research, USA \\
\textsuperscript{6}Neurology Department, New York University, USA \\
\{xj2289, sd3705, vc2558\}@columbia.edu, nima@ee.columbia.edu
}

\begin{document}
\maketitle

\footnotetext{\hspace{-20pt} \ear \brain \hspace{1pt} contribute equally.}

\begin{abstract}
Auditory foundation models, including auditory large language models (LLMs), process all sound inputs equally, independent of listener perception. However, human auditory perception is inherently selective: listeners focus on specific speakers while ignoring others in complex auditory scenes. Existing models do not incorporate this selectivity, limiting their ability to generate perception-aligned responses. To address this, we introduce Intention-Informed Auditory Scene Understanding (II-ASU) and present Auditory Attention-Driven LLM (AAD-LLM), a prototype system that integrates brain signals to infer listener attention. AAD-LLM extends an auditory LLM by incorporating intracranial electroencephalography (iEEG) recordings to decode which speaker a listener is attending to and refine responses accordingly. The model first predicts the attended speaker from neural activity, then conditions response generation on this inferred attentional state. We evaluate AAD-LLM on speaker description, speech transcription and extraction, and question answering in multitalker scenarios, with both objective and subjective ratings showing improved alignment with listener intention. By taking a first step toward intention-aware auditory AI, this work explores a new paradigm where listener perception informs machine listening, paving the way for future listener-centered auditory systems. Demo and code available\footnote{https://aad-llm.github.io}.
\end{abstract}

\begin{figure}[!t]
    \centering
    \includegraphics[width=\linewidth]{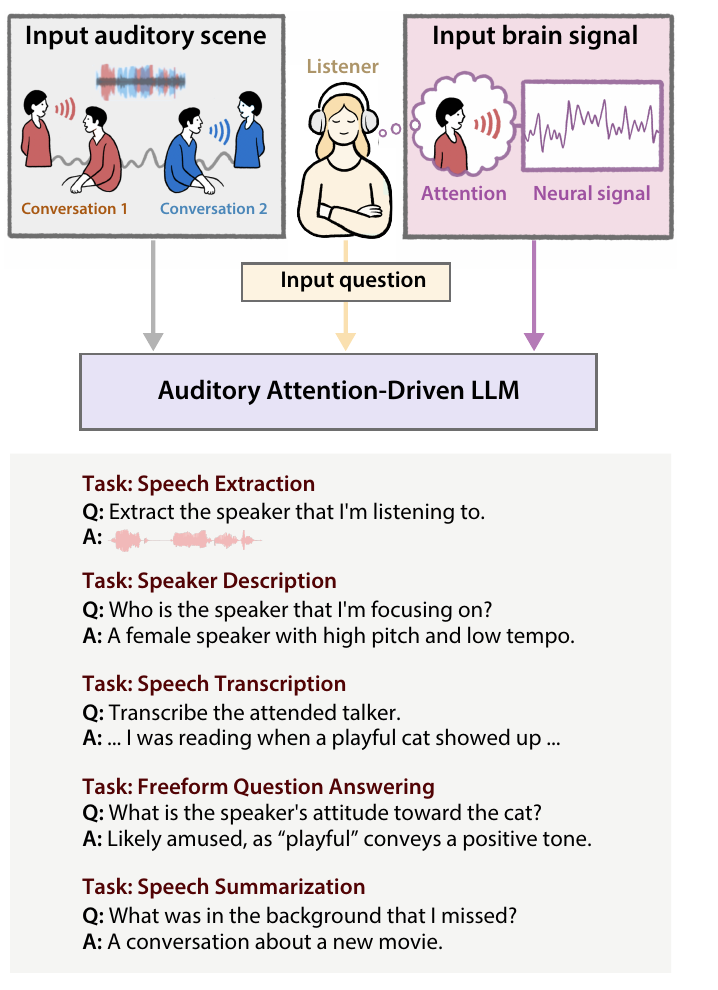}
    \caption{AAD-LLM is a brain-computer interface (BCI) for auditory scene understanding. It decodes neural signals to identify the attended speaker and integrates this information into a language model, generating responses that align with the listener’s perceptual focus.}
    \label{fig:abs}
    \vspace{-0.5cm}
\end{figure}

\section{Introduction}

The human auditory system does not process all sounds equally but selectively amplifies relevant elements while suppressing others based on the listener’s intent \cite{nima}. In a multi-speaker setting, a listener may focus on a single speaker, tune into background music, or ignore speech entirely \cite{cherry1953some, osgood1959perception, shinn2008selective}. While modern auditory foundation models, including auditory large language models (LLMs), are designed for general-purpose auditory understanding, they do not inherently account for listener intent. In applications such as assistive hearing devices, however, listener-aware processing is critical. These systems must prioritize the content most relevant to the user to improve usability in complex acoustic environments. Existing models, such as LTU \cite{ltu}, SALMONN \cite{salmonn}, and Qwen2-Audio \cite{qwen2-audio}, process all incoming audio equally, making them ineffective in scenarios where distinguishing between attended and unattended speech is essential.

Although these models excel at general auditory scene understanding \cite{MMAU, AudioBench}, they lack mechanisms to selectively process speech based on listener perception. In multi-speaker environments, they transcribe and analyze all speech sources indiscriminately, failing to separate what the user is actually attending to from background conversations \cite{wu2024just}. To address this, a listener-aware auditory AI must move beyond passive transcription and actively adapt its processing to reflect user intent.

Studies have shown that the auditory cortex encodes speech features from the attended talker \cite{nima}, allowing neural decoding methods to reconstruct or enhance the target speaker by comparing brain signal-derived representations with competing speech \cite{o2015attentional}. This line of research, also known as auditory attention decoding (AAD), has sought to infer a listener’s auditory focus from neural signals.  Both invasive \cite{cong2019aad, BISS, vishal_cong} and non-invasive methods \cite{lincoln_aad,simon_aad,eeg_aad1,pan2024neuroheed,neuroheed_plus} have been explored for attention-controlled speech extraction in hearing devices. However, AAD is primarily used for signal enhancement rather than guiding an AI’s interpretation of an auditory scene. While it could improve speech intelligibility, it does not enable models to reason about the attended content, such as summarizing speech or answering user queries based on what they perceive.

Prior studies have attempted to integrate neural signals into large language models models, incorporating neural signals to enhance multimodal processing. Some efforts have used neural data for EEG-based text generation \cite{llm_eeg1, llm_eeg2} or fMRI-informed representations \cite{llm_fmri1}, while others have integrated brain signals to improve LLM semantic understanding \cite{llm_brain_alignment1, brain_tuning}. However, these approaches focus primarily on language comprehension and semantic alignment, rather than using neural signals to refine auditory scene interpretation or speaker selection.

This paper introduces Intention-Informed Auditory Scene Understanding (II-ASU), a framework where models align their interpretation of sound with listener intent. Instead of modifying the auditory signal like AAD, we integrate attentional signals into a language model to guide how it processes and responds to auditory scenes. This approach enables reasoning beyond speech separation, allowing models to adapt responses based on listener focus rather than treating all input speech equally.
To implement this, we present Auditory Attention-Driven LLM (AAD-LLM), a prototype system that extends an auditory LLM with neural attention decoding. The model processes intracranial EEG (iEEG) to determine which speaker the listener is attending to, extracts speech representations while retaining both attended and ignored sources for contextual processing, and integrates the decoded attentional state into the LLM to generate responses that reflect listener perception.

This work makes several contributions. It introduces II-ASU as a paradigm shift from passive auditory processing to listener-aligned interpretation; It proposes AAD-LLM as the first system to integrate brain signals into an auditory LLM for attention-driven scene understanding; Finally, it evaluates AAD-LLM across multiple auditory tasks, demonstrating improved alignment with listener perception. While this study focuses on selective attention, the broader II-ASU framework could extend to other intent-detection methods, including gaze tracking, head orientation, posture shifts, or explicit user input. By incorporating such attentional signals, future intention-aware auditory AI could dynamically adapt to user perception in a wide range of applications.

\section{Intention-Informed Auditory Scene Understanding}

\subsection{Motivation and Goal}
Consider the scenario illustrated in Figure \ref{fig:abs}: both a human and a machine listener are exposed to the same auditory scene, denoted as $S$. The human listener applies selective auditory attention forming an internal representation based on their intent $I$. When asked a question $Q$, an intention-informed auditory model, such as AAD-LMM, should generate an answer $A$ that depends on $Q$, $S$, and $I$:
\begin{align}
    A = \text{MachineListener}(Q, S, I)
\end{align} \label{eq:machine}
However, existing auditory LLMs \cite{ltu, wavllm, qwen2-audio} are not intention-informed. They function as:
\begin{align}
    A^\circ = \text{AuditoryLLM}(Q, S)
\end{align} \label{eq:audio_llm}
\noindent
without any awareness of what the listener actually perceives, leading to intention-uninformed response $A^\circ$.

In real-world scenarios, the listener intention manifests in multiple ways—through facial expressions, eye gaze, head direction, verbal commands, or physical actions—but behavioral data capturing these cues remains scarce. As the first step towards intention-informed auditory scene understanding in LLMs, we focus on the most fundamental yet unexplored form of auditory intention: selective auditory attention. In this context, $I$ represents the attention of the listener to specific sound source while ignoring the other. Unlike explicit commands, this cognitive state is abstract and non-trivial to encode as input for LLMs. Our challenge, therefore, is twofold. First, decoding attentional state $I$ from brain signals $Z$ by extracting neural correlates of selective attention from intracranial EEG. Second, aligning the auditory LLM with $I$ by ensuring that responses prioritize the attended speaker, rather than treating all speakers equally.

\vspace{-0.1cm}

\subsection{Current Models in Auditory Scene Understanding}
Auditory scene understanding (ASU) is a well-studied problem, starting with knowledge and statistics-based methods \cite{ellis1996prediction, wang2006computational} to deep neural networks \cite{pann, hubert, beats}. Today, auditory LLMs represent the state-of-the-art in ASU \cite{ltu, qwen-audio}. These models take both speech and text input and generate text output. They typically consist of a speech encoder and a pretrained textual LLM, which are jointly trained for tasks including speech description, recognition, and Q\&A. Qwen2-Audio \cite{qwen2-audio}, the current state-of-the-art auditory LLM on speech understanding benchmarks \cite{MMAU, AudioBench},  serves as both a baseline and the backbone of our proposed AAD-LLM. As a brief overview, Qwen2-Audio integrates a Whisper \cite{whisper} speech encoder and a Qwen2 \cite{qwen2} LLM. The acoustic embedding of the speech input encoded by Whisper is concatenated with the textual embedding of the question, and processed together by Qwen2 to output an answer.

\begin{figure*}[h]
    \centering
    \includegraphics[width=\linewidth]{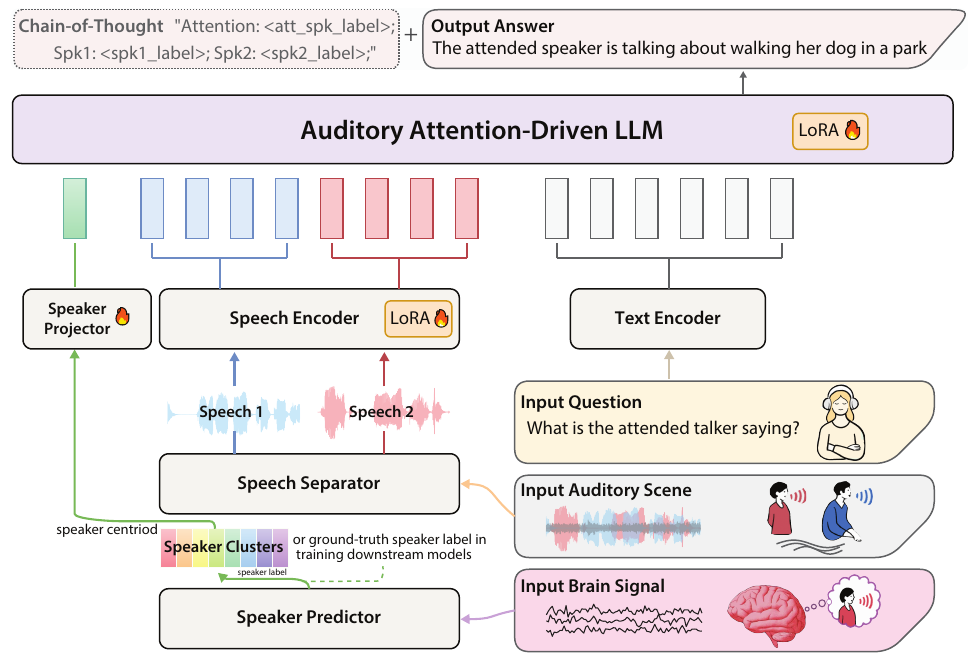}
    \caption{AAD-LLM is a multimodal, attention-aware LLM designed for auditory scene understanding. The model takes three inputs: a textual question, an auditory scene (containing mixed speech sources), and neural signals representing listener attention. Each input is processed by dedicated modules before being integrated into the LLM. AAD-LLM decodes auditory attention to determine the attended speaker and prioritizes information from the target speaker while generating responses. The model is trained to differentiate between attended and ignored speech, ensuring that its output aligns with the listener’s perceptual focus.}
    \label{fig:model}
    \vspace{-0.3cm}
\end{figure*}

\section{AAD-LLM} \label{sec:method}
We introduce AAD-LLM as a prototype system designed to address the Intention-Informed Auditory Scene Understanding (II-ASU) problem. AAD-LLM is a multimodal auditory large language model (LLM) that takes three distinct inputs: a textual question $Q$, a speech mixture $S$, and a listener’s brain signal $Z$. The model’s output $A$ is conditioned on all three inputs:
\begin{align}
    A = \text{AAD-LLM}(Q, S, Z)
\end{align} \label{eq:AAD-LLM}
\noindent
As illustrated in Figure \ref{fig:model}, AAD-LLM integrates neural attention decoding with auditory language processing. Unlike standard auditory models, AAD-LLM must process acoustic, linguistic and neural signals simultaneously. This introduces two key challenges. First, most neural recordings provide only a few minutes of data per participant, making it impractical to train a full end-to-end system jointly. Second, existing auditory attention decoding (AAD) methods \cite{simon_aad} reconstruct a temporal representation of speech from neural activity. However, these representations are continuous, noisy, and not directly compatible with discrete token-based LLMs. AAD-LLM solves these challenges by first extracting a discrete speaker identity token from brain signals, and second, using the speaker identity to condition the LLM's response generation. By decoupling brain decoding from language modeling, AAD-LLM allows intention alignment to be trained independently on large-scale speech data, while brain decoding is trained on limited neural data.

\subsection{Intention Decoding} \label{sec:decoding}

AAD-LLM introduces a speaker-based approach for decoding auditory attention from intracranial EEG signals. Instead of reconstructing speech features, we classify which speaker the listener is attending to using a discrete token representation. The process consists of two steps. First, we perform speaker clustering by applying K-means clustering to x-vectors \cite{xvector} extracted from a large corpus of thousands of speakers, ensuring no overlap with test speakers. The x-vectors, commonly used in speaker verification, are 512-dimensional embeddings, which remain frozen during training. The number of clusters K is set to 8. Next, we perform speaker prediction from neural signals. A bidirectional LSTM maps the $C$-channel $T$-length neural signal $Z \in \mathbb{R}^{C\times T} $ to a predicted cluster index $\hat{i} \in \{0, 1, ..., K-1\}$. The ground-truth label $i$ is determined by finding the closest centroid to the attended speaker’s x-vector. Further details about the model is included in Appendix \ref{app:impl}.

The intention token is represented as a speaker x-vector centroid $\hat{v} \in \mathbb{R}^D$, rather than a discrete label $\hat{i}$, to preserve numerical locality among similar speakers. A key advantage of this design is that intention decoding and intention alignment are trained separately. The speaker predictor only requires minutes of brain data, whereas the LLM can be trained on hours of speech data independently. Furthermore, this approach is modular, allowing for easy adaptation to new types of physiological signals (e.g., EEG, fNIRS, eye-gaze) without retraining the entire system.

\begin{figure*}[h]
    \centering
    \includegraphics[width=\linewidth]{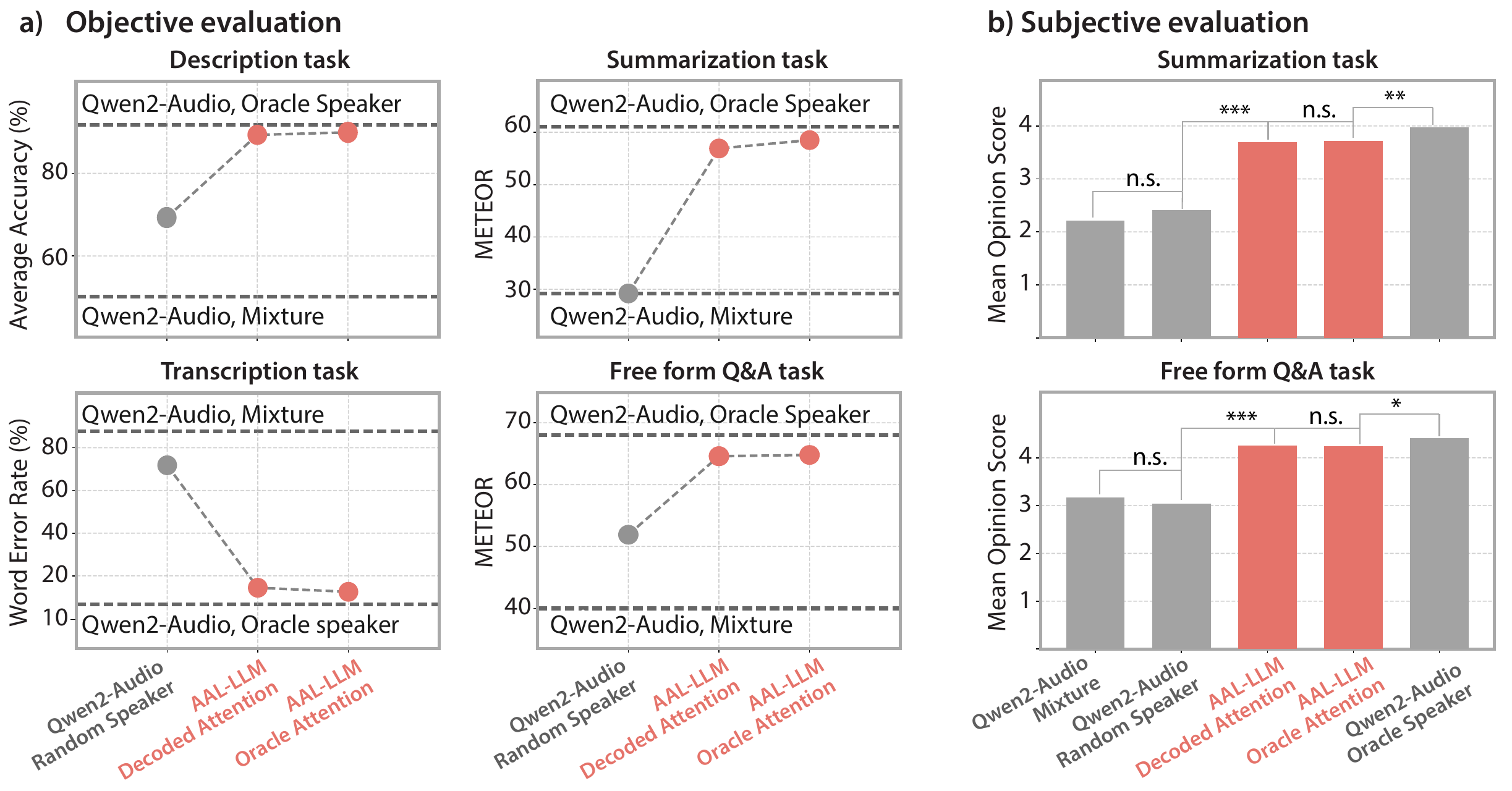}
    \caption{(a) Objective evaluation across four tasks. The dotted lines represent the lower bound (Qwen2-Audio given the mixture sound) and upper bound (Qwen2-Audio given the attended talker as oracle). ``Qwen2-Audio Random'' represents the model receiving a randomly selected talker. ``AAD-LLM Decoded Attention'' represents our method, where attention is decoded from brain signals, while "AAD-LLM Oracle Attention" represents our model given the actual attended talker as oracle. (b) Subjective evaluation measuring the alignment between model outputs and human listeners, who assessed whether the model’s response matched what should have been the answer for the attended talker. * denotes p < 0.05, ** denotes p < 0.01, *** denotes p < 0.001.}
    \label{fig:intuitive}
    \vspace{-0.3cm}
\end{figure*}

\subsection{Intention Alignment}

Once the listener’s attended speaker identity is decoded, the next challenge is to align the auditory LLM’s response with listener perception. This is achieved in three steps:

\begin{enumerate}[wide, labelwidth=!, labelindent=0pt]

\item \textbf{Embedding the intention token}:
The speaker identity token $v$ (or $\hat{v}$ during inference) is projected into the LLM’s embedding space via a linear speaker projector. It is then concatenated with the acoustic embeddings from the speech encoder and the textual embeddings from the question. Given two candidate speeches, $s_1$ and $s_2$, presented in random order, the LLM must determine which one to prioritize as the foreground and which to relegate to the background. However, simply embedding the intention into the LLM's input does not automatically enable the auditory LLM for such selective speech processing. Two more steps are needed:

\vspace{-0.5cm}
\item \textbf{Intention-informed Training}: 
Since real human attentional data is limited, we simulate attention during training by randomly assigning one of two mixed speech sources as the foreground (attended) speaker. We then assign the corresponding speaker identity token $v$ as the intention input. Finally, we train the model with intention-aware tasks such as speaker-focused transcription, selective speech summarization, and foreground/background question answering. The solution is derived from the specific speaker or speech referenced in the question, whether attended or ignored. Further details are included in Appendix \ref{app:dataset}.

\item \textbf{Chain-of-Thought Prompting}: 
Empirical experiments revealed that even after training with intention-aware answers with a large rank (512) with low-rank approximation method (LoRA) \cite{lora}, the LLM often ignored the attention token. To enforce attention usage, we introduce a structured prompt format adding chain-of-thought (CoT) \cite{cot} prefix of this form:  

\begin{tcolorbox}[
  colback=cot_color,      
  colframe=gray,       
  boxrule=0.5mm,       
  coltitle=black,      
  width=\linewidth,
  fonttitle=\bfseries, 
]
\texttt{``Attention:<att\_spk\_label>;\\Spk1:<spk1\_label>; Spk2:<spk2\_label>;"}
\end{tcolorbox}
\noindent
This CoT prompting explicitly directs the model to extract the speaker labels of the input speeches and the label of the attended speaker, all in $\{0, 1, ..., K-1\}$.  We compare with the performance of AAD-LLM trained without the CoT prefix in Appendix \ref{app:ablations}.
\end{enumerate}

\subsection{Auxiliary Module}

To further enhance speaker differentiation, AAD-LLM incorporates a speech separator based on Mamba-TasNet \cite{jiang2024speech}. This module pre-processes the speech mixture, outputting two separated streams $s_1$ and $s_2$, which are then processed by the LLM. Importantly, the speech separator is intention-uninformed, meaning that it does not inherently prioritize the attended speaker. Instead, the LLM must selectively process the correct speech stream based on brain-decoded attention signals. We compare with the performance of AAD-LLM without the separator in Appendix \ref{app:ablations}.

\subsection{Training Objective}

AAD-LLM is trained to generate a response sequence $O = \text{Cat}(\text{CoT}, A)$, combining the CoT prompt and final answer. Each output token $O_i$ is conditioned on the intention $I$ decoded from the brain, the two speech inputs $s_1$ and $s_2$, the question $Q$, and all preceding tokens $O_{1:i-1}$. This leads to the following loss function $\mathcal{L} = $
{\fontsize{10pt}{11pt}\selectfont
\begin{align*}
    -\sum_{i=2}^{N} \log P(O_i \mid \textit{SP}(I), \textit{SE}(s_1), \textit{SE}(s_2), Q, O_{1:i-1})
\end{align*}
}
where $\textit{SP}$ and $\textit{SE}$ represent the speaker projector and speech encoder, respectively, as illustrated in Figure \ref{fig:model}. Both the speech encoder and the LLM were finetuned using LoRA. The speech separator was trained to maximize the signal-to-noise ratio (SNR) of the separated speeches, and the speaker predictor was optimized using cross-entropy on ground-truth speaker labels. Additional details on training and reproducibility are provided in the Appendix \ref{app:impl}.

\begin{table}[t]
    \caption{Auditory attention decoding (AAD) accuracy and target speech extraction (TSE) performance.}
    \centering
    \begin{adjustbox}{width=\linewidth,center}
    \begin{tabular}{c|ccccc}
    \toprule
    Method & Accuracy & SNR & SI-SNR & WER$\downarrow$ & SIM \\
    \hline
    Speech Mixture & - & 0.1 & 0.1 & 37.4 & 84.4 \\
    Blind Speech Separation (BSS) & 50.4 & 5.2 & -12.3 & 64.3 & 80.6 \\
    BSS + Mel reconstruction & 92.0 & 12.0 & 9.6 & 15.2 & 94.1 \\
    BSS + Envelope reconstruction & 88.0 & 11.2 & 7.8 & 20.4 & 90.7  \\
    Target Speech Extraction & 96.0 & 12.8 & 10.4 & 14.3 & 94.8 \\ \hline
    \multicolumn{6}{c}{(Proposed Method) \textit{\textbf{Intention-Informed AAD-LLM}}} \\ \hline
    \textbf{w/ brain-decoded attention}  & 94.4 & 12.2 & 10.3 & 14.7 & 94.1 \\
    \hspace{-16mm} w/ oracle attention & 95.8 & 12.3 & 10.4 & 13.0 & 94.3 \\
    \hspace{-4mm} w/ additional clinical-15m & 97.0 & 12.6 & 11.5 & 11.4 & 94.7 \\
    \hline\hline
    Oracle Speaker (Upper Bound) & 100.0 & 13.0 & 12.8 & 8.8 & 95.5 \\
    \bottomrule
    \end{tabular}
    \end{adjustbox}
    \label{tab:aad}
    \vspace{-0.5cm}
\end{table}

\section{Experiments}

We evaluated AAD-LLM using intracranial  (iEEG) data with neural signals collected from human listeners as they attended to one conversation while another interfering conversation and background noise were present. We present only the key findings in this section (more results can be found in the appendix, including dataset curation (Appendix \ref{app:dataset}) and task specification (Appendix \ref{app:tpm}), model implementation (Appendix \ref{app:impl}), and additional results (Appendix \ref{app:results}) and analyses (Appendix \ref{app:analysis}).

\subsection{Datasets}

The iEEG \textbf{Clinical Dataset} includes six epilepsy patients implanted with intracranial electrodes as part of their medical care for epilepsy surgery. Electrode placement, clinically determined, consisted of subdural electrocorticography (ECoG) grids and/or stereo-electroencephalography (sEEG) depth electrodes. Neural signals were bandpass-filtered (0.5–30 Hz) and concatenated across participants to maximize electrode coverage. The study was approved by local IRBs, and informed consent was obtained.
Subjects listened to overlapping conversations masked with either pedestrian or babble noise. They attended to one of two simultaneous conversations, each containing two speakers taking turns. Conversations were aligned and segmented into sentences, yielding 280 training, 30 validation, and 50 testing utterances. To mitigate potential biases in the LLM’s response to speaker order, we expanded the test set by training the speaker predictor five times with different random initializations and reversing speaker order, resulting in 500 test samples.

The \textbf{Speech-Only Dataset} was collected primarily to train AAD-LLM, which requires significantly more data than the iEEG dataset. To simulate the iEEG recording conditions, we mixed two random speakers from TextrolSpeech \cite{TextrolSpeech}, a subset of LibriTTS \cite{LibriTTS}, with one background noise sample from DEMAND \cite{DEMAND}. This resulted in approximately 54,000 (85.3h) training, 1,000 (1.6h) validation, and 3,000 (4.8h) testing utterances.
Notably, there is no overlap in speakers, spoken content, or background noise between the iEEG and speech-only datasets, ensuring no information leakage during testing. Further details on both datasets are provided in Appendix \ref{app:dataset}.

\begin{table*}[t]
    \caption{Intention-informed auditory scene understanding performance when listeners were attending to one of the two speakers with background noise. A higher number indicates a better performance for all metrics except word error rate (WER). Full results with more metrics are shown in the Table \ref{tab:main_full}.}
    \begin{adjustbox}{width=\textwidth,center}
    \begin{tabular}{c|cc|cc|cc|cc}
        \toprule
        \multirow{2}{*}{\diagbox{Model}{Task}} & \multicolumn{2}{c}{Description \small{\textcolor{gray}{AVG(G, P, T)}}} & \multicolumn{2}{c}{Transcription \small{\textcolor{gray}{WER$\downarrow$}}} & \multicolumn{2}{c}{Summarization \small{\textcolor{gray}{ROUGE-L | METEOR}}
        } & \multicolumn{2}{c}{Free Q\&A \small{\textcolor{gray}{ROUGE-L | METEOR}}} \\
        & Foreground & Background & Foreground & Background & Foreground & Background & Foreground & Background  \\
        \midrule
        \multicolumn{9}{c}{(Baselines) \textit{\textbf{Auditory LLM without Intention}}} \\ \hline
        LTU-AS \cite{ltu-as} & 48.9 & 52.2 & 139.8 & 172.6 & 40.2 | 38.9 & 27.9 | 31.0 & 39.0 | 47.4  & 30.4 | 34.3 \\
        SALMONN \cite{salmonn} & 55.3 & 52.0 & 145.4  & 225.3 & 39.9 | 39.3 & 36.5 | 35.9 & 49.4 | 49.2 & 44.4 | 45.5 \\
        Qwen-Audio \cite{qwen-audio} & 36.1 & 34.3 & 82.7 & 112.8 & 29.3 | 30.0 & 24.5 | 22.8 & 35.1 | 40.1 & 30.9 | 34.3 \\
        WavLLM \cite{wavllm} & 41.7 & 37.7 & 94.7 & 128.3 & 35.6 | 34.2 & 38.3 | 38.4 & 41.5 | 40.5 & 40.2 | 38.9 \\
        GAMA \cite{gama} & 45.9 & 48.2 & n.a. & n.a. & 19.0 | 17.1 & 22.6 | 23.6 & 24.0 | 28.6  & 24.5 | 29.5 \\
        Qwen2-Audio \cite{qwen2-audio} & 50.9  & 40.1 & 90.1 & 124.7 & 27.5 | 29.0 & 15.8 | 15.9 & 39.9 | 40.4 & 34.9 | 33.3 \\
        \midrule
        \multicolumn{9}{c}{(Lower Bound) \textit{\textbf{Random Speaker + Auditory LLM}}} \\ \hline
        Qwen2-Audio \snow & 49.9 & 42.5 & 89.5 & 106.9 &  24.4 | 24.2 & 19.0 | 21.0 & 40.7 | 42.4 & 38.3 | 37.0 \\
        Qwen2-Audio \fire & 69.3 &  68.2 &  71.8 & 74.6 & 30.2 | 29.2 & 29.6 | 27.8 & 50.0 | 51.9 & 44.9 | 45.7 \\
        \midrule
        \multicolumn{9}{c}{(Proposed Baselines) \textit{\textbf{Extracted Speaker + Auditory LLM}}} \\ \hline
        Qwen2-Audio \snow & 56.2 & 44.6 & 53.6 & 73.2 &  37.0 | 41.3 & 28.2 | 33.4 & 50.8 | 54.2 & 47.9 | 46.8 \\
        Qwen2-Audio \fire & 88.1  & 77.6 & 18.5  & 24.6 & 54.5 | 53.9 & 41.0 | 40.5 & 62.3 | 65.4 & 58.0 | 60.0 \\
        \midrule
        \multicolumn{9}{c}{(Upper Bound) \textit{\textbf{Oracle Speaker + Auditory LLM}}} \\ \hline
        Qwen2-Audio \fire & 91.7 & 90.8 & 6.6 & 19.4 &  59.7 | 61.1 & 46.3 | 47.6 & 64.9 | 68.0 & 60.3 | 61.7 \\
        \midrule
        \multicolumn{9}{c}{(Proposed Method) \textit{\textbf{Intention-Informed AAD-LLM}}} \\ \hline
        \textbf{w/ brain-decoded attention} & 89.3 & 89.0 & 14.4 & 33.2 & 58.3 | 56.9 & 42.3 | 42.5 & 63.1 | 64.6 & 57.9 | 59.1 \\
        \hspace{-16mm} w/ oracle attention & 89.9 & 90.4 & 12.5 & 33.9  & 59.7 | 58.5 & 42.7 | 43.2 & 63.0 | 64.8 & 58.1 | 59.2 \\
        \hspace{-3.5mm} w/ additional clinical-15m  & 89.2 & 92.3 & \hspace{2.5mm}6.0 & 22.5 & 60.9 | 59.7 & 44.9 | 45.6  & 63.2 | 65.1  & 59.3 | 60.2 \\
        \bottomrule
    \end{tabular}
    \end{adjustbox}
    \label{tab:main}
\end{table*}

\begin{table}[t]
    \caption{Percentages of responses closer (measured by the metrics in parentheses) to the target speaker than to the other one, across different tasks for AAD-LLM with brain-decoded attention.}
    \centering
    \begin{adjustbox}{width=0.85\linewidth,center}
    \begin{tabular}{c|c|c}
        \toprule
        Task & Foreground & Background \\
        \midrule
        Description \small{\textcolor{gray}{(AVG(G, P, T))}} & 83.8 & 92.2 \\
        Transcription \small{\textcolor{gray}{(WER)}} & 94.2 & 90.4 \\
        Summarization \small{\textcolor{gray}{(ROUGE-L)}} & 92.0 & 85.4 \\
        Free Q\&A \small{\textcolor{gray}{(ROUGE-L)}} & 96.9 & 90.9 \\
        \bottomrule
    \end{tabular}
    \end{adjustbox}
    \label{tab:correctness}
    \vspace{-0.5cm}
\end{table}

\subsection{Tasks and Metrics}
We trained and evaluated AAD-LLM and other models on the following tasks and metrics:

\noindent
\textbf{Auditory Attention Decoding (AAD)} measures how accurate AAD-LLM identifies the attended speaker, a crucial step for downstream tasks. We also evaluated \textbf{Target Speech Extraction (TSE)} with signal-to-noise ratio (SNR), scale-invariant signal-to-distortion ratio (SI-SDR) \cite{sisdr}, word error rate (WER), and speaker similarity (SIM) against the clean attended speech.

We further considered four tasks for both the \textit{foreground} (attended) and \textit{background} (ignored) speaker or speech, covering different levels of speech and language processing:

\noindent
\textbf{Speaker Description} (acoustic) evaluates the accuracy of identifying the gender (G), pitch (P), and tempo (T) of the target speaker.

\noindent
\textbf{Speech Transcription} (phonetic \& syntactic) measures transcription quality using word error rate (WER) and BLEU \cite{bleu} against the target speaker's actual speech.

\noindent
\textbf{Speech Summarization} (semantic) assesses summary quality with ROUGE-L \cite{rouge} and METEOR \cite{meteor}, using three GPT-4o mini \cite{gpt4o-mini} reference summaries. The highest-scoring reference is reported.

\noindent
\textbf{Free Q\&A} (semantic \& pragmatic) evaluates responses to questions about the target speech, such as sentiment analysis, fact-checking, and named entity recognition. Three questions and reference answers were generated by GPT-4o mini, with performance measured by ROUGE-L and METEOR.

Please refer to Appendix \ref{app:tpm} for specific questions and metric computation details. While AAD-LLM was trained exclusively on these five tasks, the acquired attentional state demonstrates transferability to other tasks unseen during intention-informed training, such as \textbf{Speech Translation}. Please check Appendix \ref{app:untrained} and Table \ref{tab:untrained} for more information.

\subsection{Results}
We present the performance of AAD-LLM and other models on the iEEG clinical dataset. Results on the speech-only dataset are provided in Appendix \ref{app:offline}. By default, AAD-LLM, except for the speaker predictor, was trained on the speech-only dataset, which differs in both speakers and content from the iEEG dataset. Additionally, we report results for AAD-LLM trained with an extra 15 minutes of in-domain clinical data (``clinical-15m'').  
In all tables and figures, ``brain-decoded attention'' represents the realistic BCI use case, where the speaker predictor infers the attended speaker label from neural signals. ``Oracle attention'' serves as an upper bound, using the ground-truth speaker label from the dataset.

\subsubsection{Objective Evaluation}

Objective metrics are reported in Table \ref{tab:aad} for AAD\&TSE tasks and Table \ref{tab:main} for all other tasks. A baseline for all tasks is evaluating the speech mixture or a randomly selected speech (first six rows and the lower bound in Table \ref{tab:main}). These models, which lack attentional state, include blind speech separation and existing auditory LLMs. Their performance is close to random guessing for both foreground and background speech understanding.

Additionally, we compared AAD-LLM with standard auditory attention decoding (AAD) methods designed to separate the attended talker. We reproduced conventional AAD approaches that reconstruct the Mel spectrogram or speech envelope from brain activity to identify the attended speaker by similarity \cite{james2017, cong2019aad, simon_aad}. We also implemented an ad-hoc target speech extractor similar to \cite{BISS, pan2024neuroheed, neuroheed_plus}, optimizing SNR with the same speaker decoding method. While our model slightly underperforms the target speech extractor, it surpasses standard AAD methods and outperforms the extractor when trained with an additional 15 minutes of clinical data.

For other speech tasks, AAD-LLM outperforms all intention-uninformed auditory LLMs and a cascaded speech extractor and Qwen2-Audio (finetuned on the same data) on most metrics, particularly in transcription and summarization of the attended speaker. Notably, AAD-LLM with decoded attention performs close to the ``oracle speaker'' setting, where the target speaker is provided directly.

\begin{table}[t]
    \caption{Mean Opinion Scores (MOS) for Foreground Summarization and Free Q\&A tasks. ``Oracle speaker'' serves the performance upper bound.}
    \centering
    \begin{adjustbox}{width=\linewidth,center}
    \begin{tabular}{c|cc}
    \toprule
    Model & Summarization & Free Q\&A \\
    \hline
    Qwen2-Audio \snow & 2.21 ($\pm$ 1.44) & 3.17 ($\pm$ 1.62) \\
    Random Speaker\;+\;Qwen2-Audio \fire & 2.41 ($\pm$ 1.55) & 3.04 ($\pm$ 1.70) \\
    \hspace{0.2cm}Oracle Speaker\;+\;Qwen2-Audio \fire & 3.98 ($\pm$ 0.96) & 4.41 ($\pm$ 0.95) \\
    \hline
    \textbf{AAD-LLM w/\; brain-decoded attention}  & 3.69 ($\pm$ 1.17) & 4.25 ($\pm$ 1.13) \\
    \hspace{-1.5cm}AAD-LLM w/\; oracle attention  & 3.72 ($\pm$ 1.13) & 4.24 ($\pm$ 1.15) \\
    \bottomrule
    \end{tabular}
    \end{adjustbox}
    \label{tab:mos}
\end{table}

\subsubsection{Psychophysics Evaluation}

Subjective ratings were collected from 40 participants in psychophysics experiments replicating the auditory scenes with the same attended speakers as done in the clinical setting. Participants rated responses from five models for Summarization and Free Q\&A tasks, presented in random order, using a 5-point Likert scale. The evaluated models included AAD-LLM with brain-decoded and oracle attention, and Qwen2-Audio finetuned on single sources assessed with either a random speaker (lower bound) or the oracle speaker (upper bound). Average ratings are shown in Figure \ref{fig:intuitive} and Table \ref{tab:mos}.

Kruskal-Wallis H tests revealed significant differences in ratings between the groups for both Summarization and Free Q\&A tasks (p-values < 0.001)\footnotetext{Asterisks indicate significance levels: * p$<$0.05, ** p$<$0.01, *** p$<$0.001, and `ns' denotes non-significant results (p$\geq$0.05).}. These tests were then followed up with post-hoc pairwise Bonferroni-corrected Mann-Whitney U tests. The results of these tests show that AAD-LLM’s responses were rated significantly higher than those from the mixture and random speaker baselines across both the tasks. While mean performance increased from brain-decoded to oracle attention with ground-truth speaker labels, this improvement was not statistically significant, indicating that the neural decoding is close to its best capacity. Furthermore, both AAD-LLM models (brain-decoded and oracle attention) approach the ``oracle speaker'' upper bound, suggesting that AAD-LLM’s responses closely mirror human perception. Please see Appendix \ref{app:subj} for more details on the evaluation methods, Tables \ref{tab:stats_comparison_summ} and \ref{tab:stats_comparison_free} for p-values.

\subsubsection{Attention Validation}

To ensure AAD-LLM’s performance improvement stems from its attentional state rather than generating balanced responses for both speakers, we conducted additional analyses. Specifically, we measured the percentage of responses closer to the target speaker than the other for both foreground and background speakers across all tasks. As shown in Table \ref{tab:correctness}, over 80\% to 90\% of responses aligned more with the target speaker, confirming AAD-LLM’s effectiveness in speaker selection.

Additionally, we observed that AAD-LLM achieved similar performance in Free Q\&A tasks when using brain-decoded attention or oracle attention (Figure \ref{fig:intuitive} and Table \ref{tab:main}). Also, the Free Q\&A accuracy (96.9\%, Table \ref{tab:correctness}) surpassed the extraction accuracy (94.4\%, Table \ref{tab:aad}). These results suggest that AAD-LLM might infer the target speaker based on question content rather than attentional state, especially when speakers discuss different topics. To address this, we designed a more challenging evaluation by replacing the background speech with another speech on a similar topic that could yield a different answer. In these conditions, AAD-LLM still achieved a ROUGE-L score of 62.0 and a METEOR score of 64.2 (Table \ref{tab:tts_hard}), only slightly lower than the original scores (63.1 and 64.6), demonstrating that AAD-LLM effectively relies on attentional state to filter out the distracting speaker. More details about attention validation are included in Appendix \ref{app:tts_hard}, with other results in Appendix \ref{app:results} and analyses in Appendix \ref{app:analysis}.

\section{Conclusion}

This work introduces intention-informed auditory scene understanding (II-ASU) as a new paradigm for aligning machine listening with human perception. We present AAD-LLM, a prototype system that integrates brain signals into an auditory large language model (LLM) to decode listener attention and generate responses that align with human perception. Experimental results demonstrate that incorporating attentional state improves model performance across multiple auditory tasks, including speaker description, speech transcription, and freeform question answering. Beyond improving speech-processing capabilities, this work represents an early step toward listener-centered auditory AI, where models do not merely process sound passively but interpret auditory scenes based on what the listener perceives, which has implications for assistive hearing technologies, adaptive voice assistants, and human-computer interaction. AAD-LLM lays the groundwork for future systems that process sound in alignment with human cognitive and perceptual priorities.

\section*{Limitations}
Several limitations and challenges remain. While attention is a fundamental aspect of auditory intent, future work should explore broader cognitive signals, including task goals, semantic relevance, and perceived emotional significance of the scene. AAD-LLM also relies on intracranial EEG, which limits its current practical use, although invasive neural recordings are increasingly used more in various speech brain computer interfaces (BCI) \cite{hassan,chang1,chang2,willett2023}. While non-invasive neural recording methods such as EEG or fNIRS is desired, they presents challenges in signal quality which limits their applicability. Finally, our experiments focus on controlled two-speaker scenarios, whereas real-world auditory scenes are more complex, involving multiple speakers and environmental noise. Expanding to these settings requires further neural data collection and improved adaptation techniques.

\section*{Ethical Statement}
The development of AAD-LLM introduces exciting new possibilities for auditory scene understanding by integrating brain signals to align machine listening with human perception. This innovation has the potential to enhance communication for individuals with hearing impairments, improve virtual assistants, and advance human-computer interaction. While the model only decodes the attended speaker without accessing sensitive cognitive information, we need to remain vigilant about privacy by implementing robust safeguards and ensuring responsible data handling practices.

Within this study, approval of all ethical and experimental procedures and protocols was granted by the university's Institutional Review Board (IRB). The iEEG participants provided informed consent as per the local IRB regulations (IRB protocol number AAAD5482). The human raters evaluating model outputs also provided informed consent (IRB protocol number AAAR8655).

\section*{Acknowledgement}
This work is funded by the National Institutes of Health (NIH-NIDCD) and a grant from Marie-Josee and Henry R. Kravis.

\bibliography{custom}

\appendix

\section{Dataset Details}
\label{app:dataset}

We curated two datasets for this work. The first is a clinical dataset collected in a hospital setting from epilepsy patients with implanted intracranial electrodes. This dataset includes both neural and speech signals and was used to train the speaker predictor and evaluate AAD-LLM. The second is a synthetic speech-only dataset generated using publicly available speech and noise corpora, which was primarily used to train the AAD-LLM.

\subsection{Clinical Dataset}
\label{app:neural_dataset}

This study involved six human participants, recruited from three medical centers: two from North Shore University Hospital (NSUH), two from Columbia University Irving Medical Center (CUIMC), and two from NYU Langone Health.
All participants were undergoing clinical treatment for epilepsy and were implanted with intracranial electrodes for monitoring. 

\begin{figure}[!t]
    \centering
    \includegraphics[width=\linewidth]{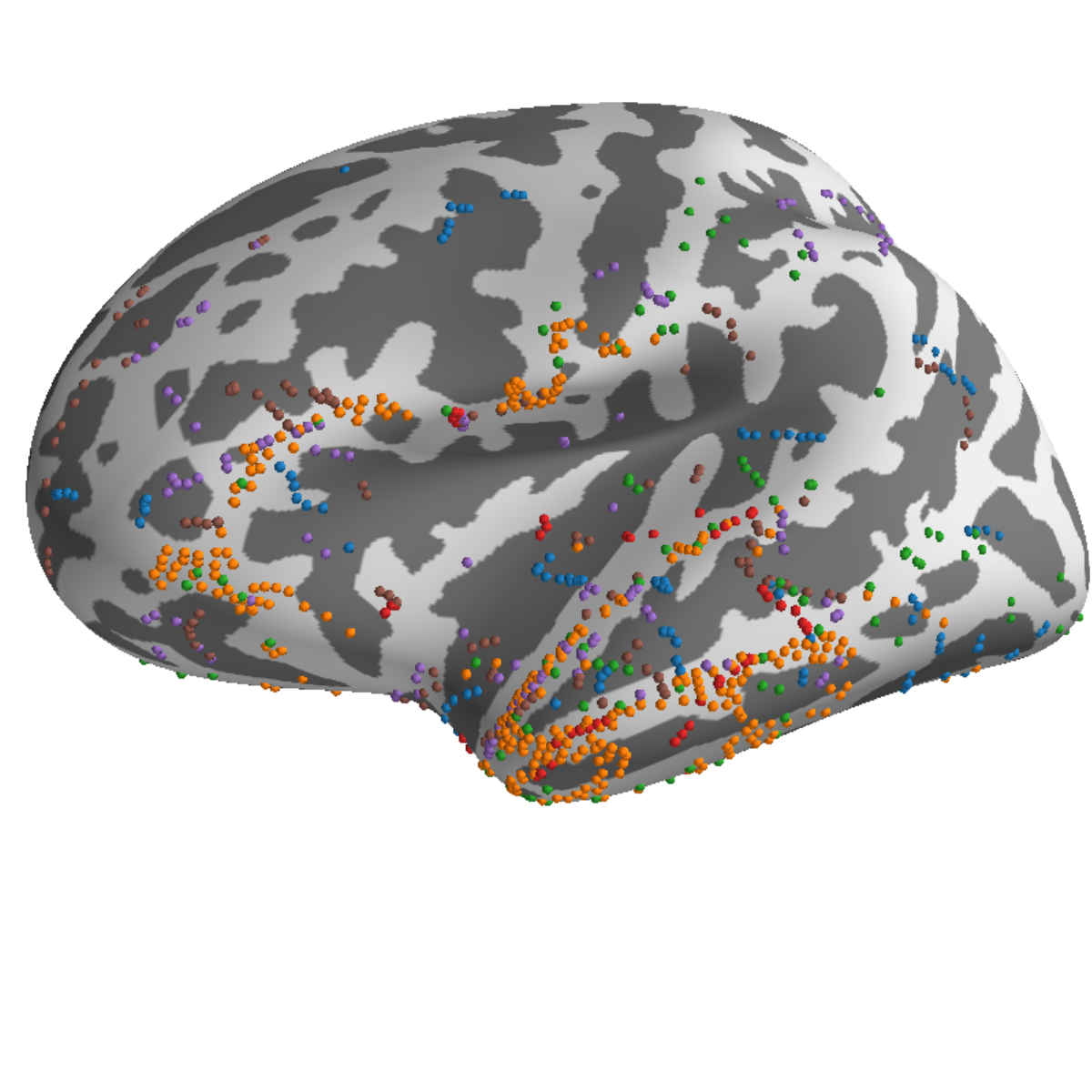}
    \caption{Electrode coverage across subjects in the clinical dataset, with each subject represented by a different color.}
    \label{fig:coverage}
\end{figure}

Each participant had electrode implants tailored to their clinical needs. Some participants were implanted with both subdural electrocorticography (ECoG) grids and stereo-electroencephalography (sEEG) depth electrodes, while others had only sEEG depth electrodes. Electrode coverage across subjects in the clinical dataset can be seen in Figure \ref{fig:coverage}, with each subject represented by a different color.

Neural recordings were bandpass-filtered to extract low-frequency components in the 0.5–30 Hz range. Electrodes that were visually identified as disconnected from anatomical tissues were excluded from the analysis. To maximize brain coverage, electrode recordings from each participant were concatenated.

All participants listened to 28 trials, with the average duration of 44.2 s (standard deviation = 2.0 s) each. The trials consisted of two concurrent and independent conversations that were equally loud and spatially separated. Diotic background noise (either “pedestrian” or “speech babble”) was also mixed along with the conversations at power either 9 or 12 dB below the power of a conversation stream. The subjects were instructed to follow (attend to) the conversation that started first. The to-be-unattended conversation started 3 seconds later. The trials were spatialized using head-related transfer functions (HRTFs) and delivered to the subjects via earphones. 

The talkers in both the conversations intentionally repeated words. To ensure that the participants tracked the target conversation, they were asked to press a button whenever they heard repeats in the target conversation. All participants were able to track the repeated words in the cued target conversation.


\begin{table*}[t]
    \caption{Intention-informed auditory scene understanding performance in the clinical testing set when listeners were attending to one of the two speakers with background noise. A higher number indicates a better performance for all metrics except word error rate (WER). This is the full version of Table \ref{tab:main}.}
    \begin{adjustbox}{width=\textwidth,center}
    \begin{tabular}{c|cc|cc|cc|cc}
        \toprule
        \multirow{2}{*}{\diagbox{Model}{Task}} & \multicolumn{2}{c}{Description \small{\textcolor{gray}{G | P | T}}} & \multicolumn{2}{c}{Transcription \small{\textcolor{gray}{WER$\downarrow$ | BLEU}}} & \multicolumn{2}{c}{Summarization \small{\textcolor{gray}{ROUGE-L | METEOR | BERT}}
        } & \multicolumn{2}{c}{Free Q\&A \small{\textcolor{gray}{ROUGE-L | METEOR | BERT}}} \\
        & Foreground & Background & Foreground & Background & Foreground & Background & Foreground & Background  \\
        \midrule
        \multicolumn{9}{c}{(Baselines) \textit{\textbf{Auditory LLM without Intention}}} \\ \hline
        LTU-AS \cite{ltu-as} & 82.6 | 38.4 | 25.8 & 76.6 | 44.0 | 36.0 & \hspace{-2.5mm} 139.8 | 44.8 & 172.6 | 7.3 & 40.2 | 38.9 | 89.7 & 27.9 | 31.0 | 87.0 & 39.0 | 47.4 | 89.9  & 30.4 | 34.3 | 88.0 \\
        SALMONN \cite{salmonn} & 82.0 | 28.0 | 56.0 & 80.0 | 32.0 | 44.0 & \hspace{-2.5mm} 145.4 | 32.6 & 225.3 | 15.6 & 39.9 | 39.3 | 88.9 & 36.5 | 35.9 |  88.4 & 49.4 | 49.2 | 91.4  & 44.4 | 45.5 | 86.8 \\
        Qwen-Audio \cite{qwen-audio} & 61.2 | 33.2 | 14.0 & \hspace{-2.5mm} 58.8 | 36.0 | 8.0 & 82.7 | 26.3 & 112.8 | 9.6 & 29.3 | 30.0 | 87.5 & 24.5 | 22.8 | 86.6 & 35.1 | 40.1 | 89.0 & 30.9 | 34.3 | 87.4 \\
        WavLLM \cite{wavllm} & 41.4 | 34.6 | 49.2 & 45.6 | 27.4 | 40.2 & 94.7 | 24.8 & 128.3 | 7.9 & 35.6 | 34.2 | 88.8 & 38.3 | 38.4 | 89.0 & 41.5 | 40.5 | 90.1 & 40.2 | 38.9 | 89.7 \\
        GAMA \cite{gama} & 64.4 | 34.6 | 38.6 & 61.8 | 37.6 | 45.2 & n.a. & n.a. & 19.0 | 17.1 | 85.1 & 22.6 | 23.6 | 85.7 & 24.0 | 28.6 | 87.2  & 24.5 | 29.5 |  86.8\\
        Qwen2-Audio \cite{qwen2-audio} & 78.6 | 37.0 | 37.0 & 67.6 | 38.8 | 14.0 & 90.1 | 30.5 & 124.7 | 11.1 & 27.5 | 29.0 | 87.3 & 15.8 | 15.9 | 85.4 & 39.9 | 40.4 | 90.0 & 34.9 | 33.3 | 88.6 \\
        \midrule
        \multicolumn{9}{c}{(Lower Bound) \textit{\textbf{Random Speaker + Auditory LLM}}} \\ \hline
        Qwen2-Audio \snow & 78.2 | 36.6 | 34.8 & 73.4 | 39.0 | 15.2 & 89.5 | 33.3 & \hspace{-2.5mm} 106.9 | 31.2 & 24.4 | 24.2 | 86.8 & 19.0 | 21.0 | 85.9 & 40.7 | 42.4 | 90.0 & 38.3 | 37.0 | 89.3 \\
        Qwen2-Audio \fire & 81.8 | 65.2 | 60.8 & 79.0 | 63.4 | 62.2 & 71.8 | 33.0 & 74.6 | 33.7 & 30.2 | 29.2 | 88.9 & 29.6 | 27.8 | 88.0 & 50.0 | 51.9 | 91.6 & 44.9 | 45.7 | 90.3 \\
        \midrule
        \multicolumn{9}{c}{(Proposed Baseline) \textit{\textbf{Extracted Speaker + Auditory LLM}}} \\ \hline
        Qwen2-Audio \snow & 91.4 | 41.2 | 36.2 & 85.2 | 33.8 | 14.8 & 53.6 | 57.1 & 73.2 | 60.3 & 37.0 | 41.3 | 89.4 & 28.2 | 33.4 | 87.7 & 50.8 | 54.2 | 91.7 & 47.9 | 46.8 | 90.5 \\
        Qwen2-Audio \fire & \hspace{-2.5mm} 100.0 | 88.6 | 75.8 & 95.4 | 59.8 | 77.6 & 18.5 | 73.0 & 24.6 | 66.2 & 54.5 | 53.9 | 93.4 & 41.0 | 40.5 | 89.7 & 62.3 | 65.4 | 93.6 & 58.0 | 60.0 | 92.5 \\
        \midrule
        \multicolumn{9}{c}{(Upper Bound) \textit{\textbf{Oracle Speaker + Auditory LLM}}} \\ \hline
        Qwen2-Audio \fire & 100.0 | 95.0 | 80.2 & 97.0 | 94.0 | 81.4 & 6.6 | 86.3 & 19.4 | 70.2 & 59.7 | 61.1 | 94.3 & 46.3 | 47.6 | 90.7 & 64.9 | 68.0 | 94.1 & 60.3 | 61.7 | 93.1 \\
        \midrule
        \multicolumn{9}{c}{(Proposed Method) \textit{\textbf{Intention-Informed AAD-LLM}}} \\ \hline
        \textbf{w/ brain-decoded attention} & 98.6 | 93.6 | 75.6 & 98.6 | 92.2 | 76.2 & 14.4 | 78.6 & 33.2 | 58.7 & 58.3 | 56.9 | 93.6 & 42.3 | 42.5 | 89.9 & 63.1 | 64.6 | 93.9 & 57.9 | 59.1 | 92.6 \\
        \hspace{-16mm} w/ oracle attention & 99.8 | 93.4 | 76.4 & 99.6 | 94.2 | 77.4 & 12.5 | 80.0 & 33.9 | 60.0 & 59.7 | 58.5 | 93.8 & 42.7 | 43.2 | 90.1 & 63.0 | 64.8 | 93.9 & 58.1 | 59.2 | 92.6 \\
        \hspace{-4mm} w/ additional Clinical-15m  & 99.6 | 87.8 | 80.2 & 99.2 | 99.0 | 78.8 & \hspace{2.5mm}6.0 | 85.9 & 22.5 | 69.2 & 60.9 | 59.7 | 94.2 & 44.9 | 45.6 | 90.8 & 63.2 | 65.1 | 93.9 & 59.3 |  60.2 | 92.9 \\
        \bottomrule
    \end{tabular}
    \end{adjustbox}
    \label{tab:main_full}
\end{table*}

\subsection{Speech-Only Dataset}
\label{app:speech_dataset}

The speech-only dataset was constructed using speech utterances from the train-clean-100 and train-clean-360 subsets of LibriTTS\footnote{LibriTTS: https://openslr.org/60} \cite{LibriTTS} along with noise samples from the DEMAND\footnote{DEMAND: https://www.kaggle.com/datasets\\/chrisfilo/demand} \cite{DEMAND}. We filtered out utterances shorter than 0.5 seconds or longer than 15 seconds and randomly combined two speech utterances of similar duration but different speakers with one of 18 environmental noise types, such as park, office, and metro station. The speech sources were normalized to have equal energy and then mixed with background noise at SNR levels of 9 or 12 dB. This process aimed to replicate the auditory conditions of the clinical dataset while introducing a more diverse set of speakers and noise types to enhance model generalizability.

In total, we generated 57,963 speeches-and-noise mixtures of 1146 speakers, which were randomly split into 53,963 (85.3 hours) for training, 1,000 (1.6 hours) for validation, and 3,000 (4.8 hours) for testing. The speech-only dataset has a distinct set of speakers and sentences from the clinical dataset. The validation and testing set were primarily used for model development and in-domain evaluation.

The gender of each speaker and the transcription of the speech were obtained from the LibriTTS corpus. Additionally, we retrieved the ``pitch'' and ``tempo'' labels for the LibriTTS utterances from TextrolSpeech\footnote{TextrolSpeech: https://github.com/jishengpeng/TextrolSpeech} \cite{TextrolSpeech}. Both pitch and tempo were quantized into three levels. For pitch, utterances with a fundamental frequency below 136.6 Hz were labeled as ``low'', above 196.1 Hz as ``high'', and those in between as ``normal''. For tempo, utterances with an average speaking rate slower than 0.39 seconds per word were labeled as ``low'', faster than 0.25 seconds per word as ``high'', and those in between as ``normal''. We applied these same pitch and tempo thresholds to annotate speech utterances in the clinical dataset, serving as the ground-truth labels for the speaker description task.

\section{Model and Training Details}
\label{app:impl}

\subsection{Auditory Large Language Model}

AAD-LLM adopts the backbone of Qwen2-Audio \cite{qwen2-audio}. The pretrained checkpoint Qwen2-Audio-7B-Instruct is publicly available\footnote{Qwen2-Audio: https://huggingface.co\\/Qwen/Qwen2-Audio-7B-Instruct}. We further finetuned both the LLM and the speech encoder on the speech-only dataset using low-rank approximation, LoRA \cite{lora}, on the key, query, and value matrices of attention layers and the weight matrices of multilayer perceptrons. We used a rank of 512, an $\alpha$ of 512, and a dropout of 0.05 by default, which adds around 16.5\% of trainable parameters.

We added a special token <ATT> as the placeholder for the listener's attention (speaker vector). In reality, <ATT> is <|extra\_124|> from Qwen2-Audio's reserved special token sets. Then, the entire input to the LLM looks like the following:  

\begin{tcolorbox}[
  colback=gray!10,      
  colframe=gray,        
  boxrule=0.5mm,        
  coltitle=black,       
  width=\linewidth,
  fonttitle=\bfseries,  
]
\texttt{system: You are a helpful assistant. \\
user: Attention: <ATT> \\
Audio 1: <speech1> \\
Audio 2: <speech2> \\
Question: <question> \\
Solution: \textcolor{pink}{Attention:<att\_spk\_label>;\\Spk1:<spk1\_label>; Spk2:<spk2\_label>;}\\
\textcolor{red!70}{<solution>}
}
\end{tcolorbox}
<att\_spk\_label>, <spk1\_label>, and <spk2\_label> are ground-truth labels of the attended speaker, the first input speaker, and the second input speaker. All are integers from 0 to K-1, with K=8 by default. (<att\_spk\_label> $\neq$ <ATT>. The former is an integer; The latter is the projected speaker vector.)
The tokens of pink and red parts corresponding to the chain-of-thought prefix and the actual solution were optimized. The maximum number of allowed tokens is 1024.

\noindent
\textbf{Speaker Projector} is a linear layer from 512 to 4096, from the dimension of x-vector to the dimension of the LLM. It was optimized jointly with the audio encoder and the LLM. The projected speaker vector replaces the embedding at the <ATT> token.

We trained the model using an AdamW optimizer \cite{adamw} with a learning rate of \( 1 \times 10^{-4} \), scheduled with a cosine decay and a warmup ratio of 10\% of total steps. The model was trained for 20 epochs using a batch size of 1 and gradient accumulation steps of 8 for an effective batch size of 8. Mixed precision training with bf16 is enabled to optimize memory usage and training speed. The total training took around 4.5 days in an NVIDIA L40 GPU.

We trained all ablation models and baseline Qwen2-Audio on a single speaker with the exact same training configuration.

\subsection{Speaker Clusters}

K=8 speaker clusters were obtained by K-Means. We randomly sampled 10,000 utterances of 1137 speakers from LibriTTS and extracted their x-vectors using a pretrained speaker verification model. The x-vector extractor\footnote{TDNN for x-vector extraction: https://huggingface.co\\/speechbrain/spkrec-xvect-voxceleb} has an output dimension of 512 and was kept frozen. Please see the right subfigure of Figure \ref{fig:xvectors} for the distribution of 10,000 x-vectors colored by the nearest cluster.

\subsection{Speaker Predictor} \label{app:spk_pred}

The speaker predictor classifies the neural signal \( Z \) into one of \( K = 8\) discrete speaker labels. The predictor consists of a bidirectional recurrent neural network (RNN) followed by a temporal pooling layer and a fully connected classification head.

Before passing the neural signals into the recurrent layer, layer normalization is applied. The recurrent module is a bidirectional long short-term memory (LSTM) network with a hidden state size of \( S = 64 \), resulting in an output dimension of \( 2S = 128 \) due to bidirectionality. The LSTM output is processed by a mean pooling layer, followed by a fully connected layer with 128 hidden units and a final softmax activation for classification.

We trained the predictor on 280 samples with aligned neural and speech signals in the clinical training set. We obtained the ground-truth speaker label from the speech signal. The predictor was then optimized by cross-entropy loss, with the predicted speaker labels against ground-truth labels. We optimized the predictor with an Adam optimizer of a constant learning rate of \( 1 \times 10^{-4} \), with a batch size of 1, for a total of 30 epochs, which took fewer than 10 minutes on an GPU.

\subsection{Speech Separator}

The speech separator reproduces Mamba-TasNet (M) \cite{jiang2024dual, jiang2024speech} which has a linear waveform encoder \& decoder and 32 Mamba \cite{mamba} layers with a dimension of 256. The model is tiny (15.6M parameters) and intention-uninformed. It was trained separately on the speech-only dataset. The model was trained using an Adam optimizer \cite{adam} with a learning rate of \( 2 \times 10^{-4} \). A cosine learning rate schedule with 20,000 warmup steps and a ReduceLROnPlateau scheduler was applied, halving the learning rate if performance plateaus after 30 epochs. The model was trained for 50 epochs with a batch size of 4 and gradient clipping set to 5. Mixed precision training with bf16 is enabled for efficiency. The training was conducted on an NVIDIA L40 GPU.

The target speech extractor, as the baseline model in Tables \ref{tab:aad} and \ref{tab:main}, matches the speech separator in both the number of Mamba layers and the dimension, ensuring comparable acoustic processing capabilities. To condition the extractor on the speaker, we fused the acoustic features with the speaker vector following the approach in \cite{jiang2024listen}. The extractor was trained on the same data with an identical training configuration.

\section{Tasks, Prompts, and Metrics}
\label{app:tpm}

\subsection{Questions and Solutions}
We wrote eight different questions for each task except Free Q\&A, for which GPT-4o mini generates three questions and reference solutions uniquely for each utterance.

Here are three example questions for foreground speaker description:

\begin{verbatim} 
Q1: "Describe the attended speaker." 
Q2: "Please write a description of 
the attended speaker." 
Q3: "Can you identify the person the 
subject is listening to?" \end{verbatim}

The solution to the description question is formatted as:

\begin{verbatim} 
A: "A <gender_label> speaker 
with <pitch_label> pitch 
and <tempo_label> tempo."
\end{verbatim}

The pitch and tempo labels are ``low'', ``normal'', or ``high'' (Appendix \ref{app:speech_dataset}).

Here are three example questions for background speech summarization:

\begin{verbatim} 
Q1: "What is the background speaker 
talking about?",
Q2: "Can you summarize the speech 
of the speaker being ignored?",
Q3: "What topic is the background 
speaker discussing?",
\end{verbatim}

We gave GPT-4o mini the transcription of the speech to generate three candidate summaries as the solutions.

In each training epoch, a random question from a random speaker was sampled for every utterance.

\subsection{Free Q\&A Generation}

Three pairs of freeform questions and answers were generated for either speaker of each utterance with the following prompt to GPT-4o mini:
\begin{verbatim} 
"You are listening to a conversation. 
One speaker said: "<transcription>" 
Raise 3 questions and provide an answer for 
each one of them. Be short. Don't use any 
information outside the speech.
Answer in this format:
Question 1: ... Answer 1: ...
Question 1: ... Answer 2: ...
Question 1: ... Answer 2: ..."
\end{verbatim}

Here are three example pairs of free Q\%A questions and solutions from three different utterances and speakers:

\begin{verbatim} 
Q1: "How might the description affect the 
mood of the conversation?" 
A1: "It may create a sense of confusion 
or disorientation due to the winding 
nature of the street." 
Q2: "Why are they referred to as 
'The Flying Stars'" 
A2: "Because they've been stolen 
so often."
Q3: "What is the name of the establishment 
mentioned?" 
A3: "The name of the establishment 
mentioned is the Royal Oak." 
\end{verbatim}

\subsection{Evaluation Details}

Most of the evaluation metrics were computed by Huggingface's Evaluate package\footnote{https://huggingface.co/docs/evaluate}. In particular:
\begin{verbatim} 
bleu_metric = evaluate.load("bleu")
rouge_metric = evaluate.load("rouge")
meteor_metric = evaluate.load("meteor")
bert_metric = evaluate.load(
    "bertscore", lang='en'
)
\end{verbatim}
Prior to computing these metrics, we applied text preprocessing steps to ensure a fair comparison, particularly benefiting the baseline models. The preprocessing included:

\begin{itemize}
    \item Remove common prefixes like ``The attended speaking is discussing about'' and ``Spoken text: ''.
    \item Apply BasicTextNormalizer\footnote{from transformers.models.whisper.english\_normalizer}.
    \item For speech translation (English$\rightarrow$Chinese), apply jieba tokenizer on Chinese\footnote{https://github.com/fxsjy/jieba}.
\end{itemize}

Additionally, for the target speech extraction task (Table \ref{tab:aad}), the word error rate (WER) was computed by speech recognition model Whisper \cite{whisper} (whisper-large-v3\footnote{https://huggingface.co/openai/whisper-large-v3}) and the speaker similarity SIM was computed by WavLM \cite{wavlm} (wavlm-base-plus-sv\footnote{https://huggingface.co/microsoft/wavlm-base-plus-sv}) for speaker verification.

For the speaker description task (Table \ref{tab:main}), since baseline pretraind models were not trained to output all of the gender, pitch, and tempo in one response, we evaluated them separately as three simpler tasks. We also informed the models of the pitch and tempo cutoff corresponding to ``low'', ``normal'', and ``high'' in the question.
 
\section{Additional Results}
\label{app:results}

\begin{table}[t]
    \caption{AAD-LLM's attentional state can be transferred to other tasks unseen in intention-informed training, such as speech translation.}
    \begin{adjustbox}{width=\columnwidth,center}
    \begin{tabular}{c|cc}
        \toprule
        \multirow{2}{*}{\diagbox{Model}{Task}} & \multicolumn{2}{c}{Translation \small{\textcolor{gray}{BLEU | METEOR}}}  \\
        & Foreground & Background \\
        \midrule\
        Qwen2-Audio \snow & 41.0 | 28.6 & 34.2 | 23.4  \\
        Random Speaker\;+\;Qwen2-Audio \fire & 40.5 | 26.5  & 22.1 | 9.8 \\
        \hspace{0.2cm}Oracle Speaker\;+\;Qwen2-Audio \fire & 66.8 | 57.5 & 51.9 | 43.4 \\ \hline
        \textbf{AAD-LLM w/ brain-decoded attention} & 61.8 | 53.8  & 45.5 | 35.1 \\
        \bottomrule
    \end{tabular}
    \end{adjustbox}
    \label{tab:untrained}
\end{table}

\subsection{Evaluation on Untrained Task: Speech Translation}
\label{app:untrained}

We further investigated if the selective processing ability of the foreground or the background speech was restricted to the tasks AAD-LLM was trained on or if this attentional state could generalize to new tasks. To test this, we evaluated the English-to-Chinese speech translation performance of AAD-LLM after intention-informed training on other tasks. The results, presented in Table \ref{tab:untrained}, were compared against oracle Chinese transcriptions generated by GPT-4o \cite{gpt4o} from the corresponding English speech.

The results show that AAD-LLM significantly outperforms Qwen2-Audio on both speech mixtures and random speakers, while approaching the upper bound set by Qwen2-Audio with oracle speakers—a pattern consistent with the results observed in the tasks on which AAD-LLM was trained. This suggests that auditory attention is a generalizable capability that can be successfully transferred to other tasks without further training.

\subsection{Evaluation on Speakers Talking About the Same Topic}
\label{app:tts_hard}

The performance deviation on Free Q\&A tasks is closer across models with and without attention compared to other tasks (Table \ref{tab:main}). This suggests a possibility that some questions can be answered solely based on the content of the question and the two candidate speech inputs by selecting the only speech that is related to the question without the need to decode the listener's attention. To verify that AAD-LLM genuinely relies on attention rather than cheating on content, we designed a more challenging Free Q\&A testing set.

We replaced the background speech in the clinical testing set with another speech on the same topic but providing a different answer to the question than the foreground speech. The content of the replacement is generated by GPT-4o \cite{gpt4o} with the following prompt:

\begin{verbatim} 
Foreground Speech: "<transcription>"
Question: "<question>"
Solution: "<solution>"

I want to test whether a QA system can 
distinguish between the foreground and 
background speakers. According to the 
foreground speech, the question, and 
the solution given, I want you to come
up with a background speech in words. 
The background speech can also answer 
the sentence, but the solution should 
be different or even opposite. Try to 
have roughly the same number of words 
as the foreground speech.

Give me the sentence of the background 
speech directly. No explanation is needed.
\end{verbatim}

\begin{table}[t]
    \caption{Free Q\&A on same-topic speech mixtures. Both foreground and background speech can be used to answer the question, so auditory attention to the correct speaker is necessary to answer the question correctly.}
    \begin{adjustbox}{width=\columnwidth,center}
    \begin{tabular}{c|c}
        \toprule
        \diagbox{Model}{Free Q\&A} & \multicolumn{1}{c}{\textcolor{gray}{ROUGE-L | METEOR | BERT}} \\
        \midrule\
        Qwen2-Audio \snow &  41.6 | 44.1 | 90.3 \\
        Random Speaker\;+\;Qwen2-Audio \fire & 49.9 | 51.4 | 91.9 \\
        \hspace{0.2cm}Oracle Speaker\;+\;Qwen2-Audio \fire & 65.4 | 68.3 | 94.2 \\ \hline
        \textbf{AAD-LLM w/ brain-decoded attention} & 62.0 | 64.2 | 93.7 \\
        \bottomrule
    \end{tabular}
    \end{adjustbox}
    \label{tab:tts_hard}
\end{table}

\begin{table*}[t]
    \caption{Offline evaluation on speech-only data. One of the two speaker is randomly designated as the foreground and the other as the background.}
    \begin{adjustbox}{width=\textwidth,center}
    \begin{tabular}{c|cc|cc|cc|cc}
        \toprule
        \multirow{2}{*}{\diagbox{Model}{Task}} & \multicolumn{2}{c}{Description \small{\textcolor{gray}{G | P | T}}} & \multicolumn{2}{c}{Transcription \small{\textcolor{gray}{WER$\downarrow$ | BLEU}}} & \multicolumn{2}{c}{Summarization \small{\textcolor{gray}{ROUGE-L | METEOR | BERT}}
        } & \multicolumn{2}{c}{Free Q\&A \small{\textcolor{gray}{ROUGE-L | METEOR | BERT}}} \\
        & Foreground & Background & Foreground & Background & Foreground & Background & Foreground & Background  \\
        \midrule
        \multicolumn{9}{c}{(Baselines) \textit{\textbf{Auditory LLM without Intention}}} \\ \hline
        LTU-AS \cite{ltu-as} & 73.7 | 47.1 | 26.7 & 73.1 | 46.5 | 28.1 & 148.0 | 26.3 & 154.3 | 26.9 & 19.5 | 21.6 | 85.8 & 17.6 | 19.2 | 85.4 & 33.1 | 37.8 | 88.5 & 32.6 | 36.9 | 88.4 \\
        SALMONN \cite{salmonn} & 73.2 | 33.5 | 71.4 & 76.5 | 34.6 | 71.5 & 122.1 | 25.5 & 118.5 | 27.4 & 16.4 | 18.0 | 84.8 & 15.4 | 16.8 | 84.6 & 44.5 | 46.5 | 90.1 & 43.6 | 45.5 | 89.9 \\
        Qwen-Audio \cite{qwen-audio} & 48.2 | 37.5 | 8.5 & 47.6 | 37.6 | 9.0 & 96.0 | 18.5 & 98.1 | 19.5 & 17.5 | 18.1 | 85.1 & 17.8 | 16.4 | 85.1 & 31.4 | 36.3 | 87.5 & 30.7 | 35.5 | 87.5 \\
        WavLLM \cite{wavllm} & 53.6 | 31.9 | 53.5 & 54.0 | 33.7 | 51.2 & 84.4 | 21.0  & 82.8 | 20.7 & 20.6 | 19.4 | 86.1 & 19.4 | 18.9 | 85.9 & 39.7 | 38.1 | 89.2 & 39.1 | 37.4 | 89.1 \\
        GAMA \cite{gama} & 60.2 | 32.6 | 52.9 & 58.8 | 34.5 | 50.5 & n.a. & n.a. & 17.4 | 15.2 | 84.7 & 18.6 | 16.9 | 84.8 & 23.8 | 28.5 | 86.8 & 23.4 | 27.9 | 86.7 \\
        Qwen2-Audio \cite{qwen2-audio} & 71.9 | 34.4 | 59.7 & 74.3 | 34.2 | 40.1 & 81.0 | 25.7  & 80.4 | 27.9  & 18.5 | 20.4 | 85.5 & 17.4 | 18.9 | 85.3 & 41.5 | 42.4 | 89.6 & 40.2 | 41.3 | 89.4 \\
        \midrule
        \multicolumn{9}{c}{(Lower Bound) \textit{\textbf{Random Speaker + Auditory LLM}}} \\ \hline
        Qwen2-Audio \fire & 75.9 | 63.4 | 73.1 & 72.6 | 63.1 | 73.8 & 56.8 | 45.2 & 57.7 | 45.2 & 27.4 | 25.7 | 88.3 & 28.4 | 26.9 | 87.8 & 47.1 | 48.3 | 90.7 & 46.9 | 48.3 | 90.7 \\
        \midrule
        \multicolumn{9}{c}{(Proposed Baselines) \textit{\textbf{Extracted Speaker + Auditory LLM}}} \\ \hline
        Qwen2-Audio \fire & 99.1 | 89.5 | 87.4 & 95.4 | 82.7 | 84.6 & 14.0 | 82.3 & 14.3 | 82.1 & 44.1 | 42.7 | 91.2 & 41.8 | 41.6 | 90.3 & 51.5 | 52.9 | 91.6 & 51.2 | 52.7 | 91.6 \\
        \midrule
        \multicolumn{9}{c}{(Upper Bound) \textit{\textbf{Oracle Speaker + Auditory LLM}}} \\ \hline
        Qwen2-Audio \fire & 99.7 | 94.6 | 91.6 & 96.4 | 91.7 | 90.0 & 1.9 | 94.8 & 2.2 | 94.6 & 49.7 | 49.0 | 92.2 & 45.6 | 45.9 | 91.0 & 65.9 | 67.9 | 94.2 & 64.9 | 67.3 | 94.0 \\
        \midrule
        \multicolumn{9}{c}{(Proposed Method) \textit{\textbf{Intention-Informed AAD-LLM}}} \\ \hline
        \hspace{-6.5mm} w/ oracle attention & 99.4 | 90.9 | 88.0 & 99.5 | 92.2 | 87.7 & 10.6 | 86.3 & 10.8 | 86.1 & 46.4 | 45.2 | 91.6 & 46.3 | 45.4 | 91.5 & 64.2 | 65.7 | 93.8 & 63.1 | 65.0 | 93.7 \\
        \bottomrule
    \end{tabular}
    \end{adjustbox}
    \label{tab:offline}
\end{table*}

\begin{table*}[t]
    \caption{Ablations of AAD-LLM.}
    \begin{adjustbox}{width=\textwidth,center}
    \begin{tabular}{c|cc|cc|cc|cc}
        \toprule
        \multirow{2}{*}{\diagbox{AAD-LLM}{Task}} & \multicolumn{2}{c}{Description \small{(G | P | T)}} & \multicolumn{2}{c}{Transcription \small{(WER$\downarrow$ | BLEU)}} & \multicolumn{2}{c}{Summarization \small{(ROUGE-L | METEOR | BERT)}} & \multicolumn{2}{c}{Free Q\&A \small{(ROUGE-L | METEOR | BERT)}} \\
        & Attended & Unattended & Attended & Unattended & Attended & Unattended & Attended & Unattended  \\
        \midrule
        \multicolumn{9}{c}{\textit{\textbf{Rank of LoRA}}} \\
        \hline
        r=32 & 95.2 | 80.8 | 71.6 & 97.0 | 74.2 | 75.6 & 20.7 | 73.3 & 33.4 | 60.1 & 55.3 | 53.5 | 92.7 & 40.1 | 40.4 | 89.6 & 62.7 | 64.9 | 93.7 & 56.6 | 58.8 | 92.5 \\
        r=128 & 97.6 | 83.0 | 77.6 & 98.0 | 82.2 | 80.8 & 16.3 | 75.0 & 34.0 | 59.6 & 57.1 | 56.5 | 93.4 & 43.8 | 44.2 | 90.4 & 62.2 | 64.5 | 93.7 & 58.0 | 60.4 | 92.7 \\
        \hline
        \multicolumn{9}{c}{\textit{\textbf{Necessity of Components}}} \\
        \hline
        without CoT & 97.8 | 86.8 | 72.0 & 98.8 | 79.2 | 58.8 & 34.9 | 65.2 & \hspace{-2.5mm} 100.6 | 15.8 & 38.3 | 37.0 | 90.2 & 25.6 | 24.8 | 87.6 & 63.8 | 64.8 | 93.9 & 59.0 | 59.7 | 92.8 \\
        without Separation & 99.2 | 77.6 | 61.6 & 99.2 | 83.0 | 70.4 & 41.8 | 50.9 & 73.7 | 29.0 & 40.5 | 38.9 | 90.6 & 29.2 | 27.8 | 87.8 & 59.6 | 61.3 | 93.1 & 53.2 | 54.2 | 91.7 \\
        \hline
        \multicolumn{9}{c}{\textit{\textbf{Types of Features and Numbers of Cluster for Attention Decoding}}} \\
        \hline
        xvect-4 & 96.2 | 87.2 | 71.8 & 96.2 | 88.2 | 74.8 & 37.0 | 60.8 & 57.7 | 47.6 & 46.3 | 45.2 | 91.6 & 34.7 | 34.4 | 88.9 & 64.5 | 66.3 | 94.1 & 58.9 | 60.7 | 92.8 \\
        xvect-16 & 97.0 | 84.2 | 67.0 & 97.4 | 87.0 | 76.8 & 31.5 | 65.7 & 43.0 | 53.1 & 51.2 | 49.9 | 92.2 & 38.7 | 38.7 | 89.6 & 61.0 | 62.9 | 93.5 & 57.2 | 59.2 | 92.4 \\
        xvect-32 & 94.6 | 82.2 | 70.0 & 93.8 | 83.8 | 71.2 & 33.4 | 66.8 & 48.1 | 50.8 & 50.5 | 49.6 | 92.3 & 36.7 | 36.1 | 89.4 & 62.4 | 64.2 | 93.7 & 58.6 | 61.3 | 92.6 \\
        style-8 & 93.0 | 72.4 | 70.2 & 94.4 | 78.2 | 72.2 & 44.4 | 56.6 & 51.6 | 46.7 & 46.0 | 44.6 | 91.2 & 32.1 | 31.7 | 88.5 & 62.6 | 64.9 | 93.7 & 58.2 | 60.3 | 92.6 \\
        random-8 & 79.6 | 64.4 | 61.8 & 83.2 | 67.0 | 68.6 & 63.1 | 43.6 & 74.8 | 30.7 & 36.3 | 34.5 | 89.6 & 28.8 | 27.9 | 88.1 & 63.7 | 66.0 | 94.0 & 57.2 | 59.0 | 92.5 \\
        \hline
        \multicolumn{9}{c}{\textit{\textbf{Default: }r=512, with CoT and Separation, xvect-8, without in-domain data}} \\
        \hline
        AAD-LLM & 98.6 | 93.6 | 75.6 & 98.6 | 92.2 | 76.2 & 14.4 | 78.6 & 33.2 | 58.7 & 58.3 | 56.9 | 93.6 & 42.3 | 42.5 | 89.9 & 63.1 | 64.6 | 93.9 & 57.9 | 59.1 | 92.6 \\
        \bottomrule
    \end{tabular}
    \end{adjustbox}
    \label{tab:ablation}
\end{table*}

Then, we synthesized one of four speakers (two females and two males) by KokoroTTS\footnote{https://huggingface.co/hexgrad/Kokoro-82M; The four speakers are ``'af\_bella'', ``af\_sarah'', ``am\_adam'', and ``am\_michael''.}, a human-level text-to-speech model based on StyleTTS 2's architecture \cite{styletts2}, given the generated text. We then mixed the new background speech with the original foreground speech and the background noise. We asked the same question again on this same-topic dataset. The results are shown in Table \ref{tab:tts_hard}, averaged across four new speakers.

In this more challenging dataset with more distraction in the spoken content, AAD-LLM still achieved a ROUGE-L score of 62.0 and a METEOR score of 64.2, only slightly below the original scores of 63.1 and 64.6. This means that AAD-LLM is using its attentional state to filter out the distracting speaker rather than using the content.

\subsection{Evaluation on Speech-only Dataset}
\label{app:offline}

We also evaluated AAD-LLM on a speech-only dataset (LibriTTS and DEMAND mixtures) described in Appendix \ref{app:speech_dataset}. One of the two speakers was randomly designated as the foreground, and models were tested on speaker description, speech transcription, speech summarization, and free Q\&A. We used oracle attention (ground truth labels) for evaluation since the brain signal is not available for this dataset.

AAD-LLM achieved high classification accuracy in speaker description (99.4\% gender, 90.9\% pitch, 88.0\% tempo), outperforming Qwen2-Audio and other baseline models. In transcription, AAD-LLM significantly lowered the WER to 10.6\% compared to Qwen2-Audio’s 81.0\%. Summarization results showed a ROUGE-L score of 46.4, exceeding baseline models. In Q\&A, AAD-LLM achieved a ROUGE-L of 64.2 and METEOR of 65.7, surpassing Qwen2-Audio’s 41.5 and 42.4.

The performance of AAD-LLM on the speech-only dataset is comparable to that of the clinical dataset. The model consistently outperformed other auditory LLMs, demonstrating its advantage in processing multi-speaker scenarios. These findings also highlight AAD-LLM’s potential as an effective offline system for non-BCI applications, provided the x-vector of the target speaker is known.

\begin{figure*}[h]
    \centering
    \includegraphics[width=\linewidth]{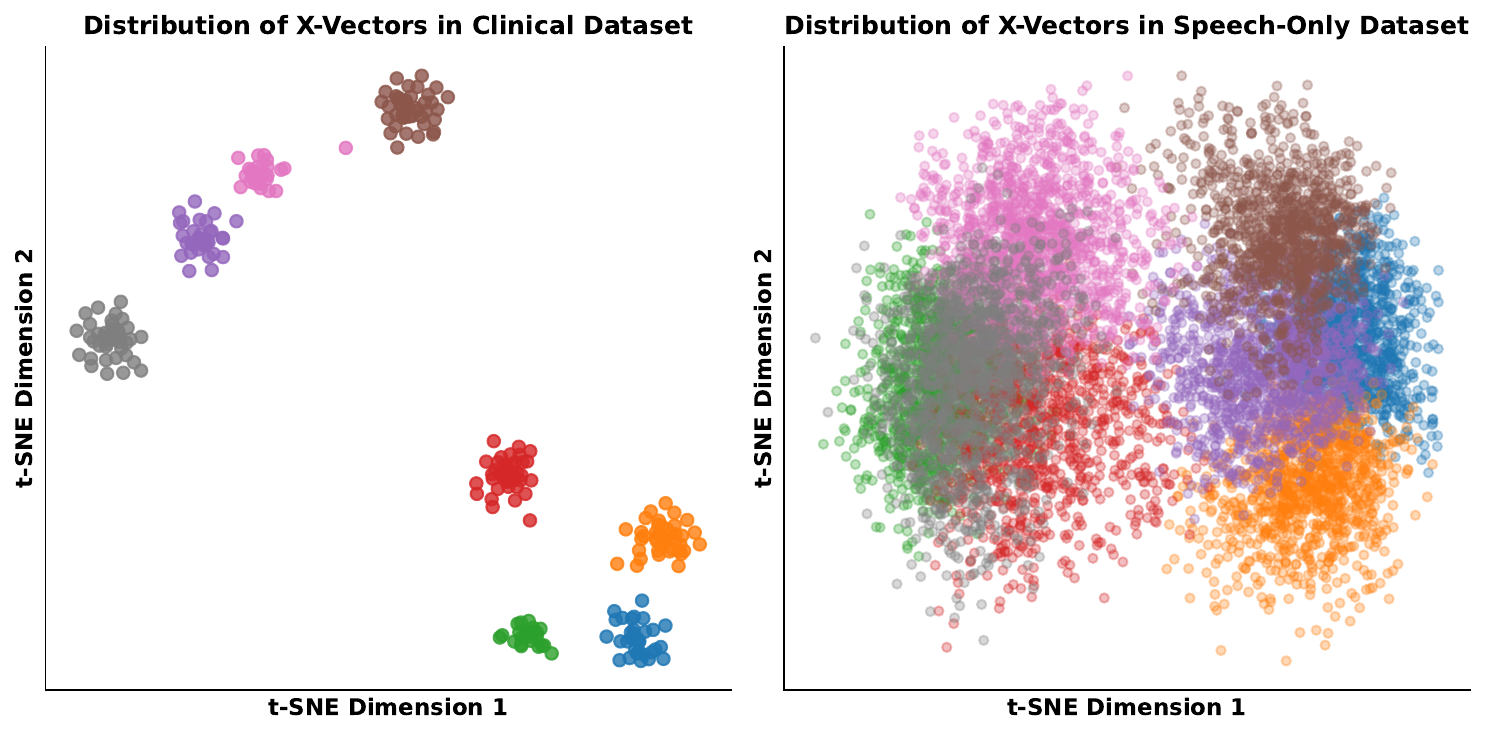}
    \caption{t-SNE visualization of x-vectors after K-Means clustering (K=8). The left plot (clinical dataset, 280 sentences) shows distinct clusters for eight speakers. The right plot (speech-only dataset, 10,000 sentences) reveals two main clusters for male and female speakers. Colors indicate K-means clustering with eight groups.}
    \label{fig:xvectors}
\end{figure*}

\subsection{Impact of Background Noise Types}
\label{app:noise}

The clinical test set contains speech mixtures under two distinct background noises: babble and pedestrian. The performance with respect to the noise type is shown in Table~\ref{tab:noise}. Transcription error (WER) and summarization quality (ROUGE-L \& METEOR) are consistently better under babble noise compared to pedestrian noise. These results suggest that pedestrian noise, characterized by unpredictable acoustic dynamics, presents greater difficulty for the AAD-LLM, slightly degrading the recognition of attended speech and subsequent response generation.

\begin{table*}[htbp]
  \centering
  \caption{Performance across noise types.}
  \label{tab:noise}
  \resizebox{\textwidth}{!}{%
    \begin{tabular}{lcccccccc}
      \toprule
      \textbf{NoiseType}
        & \textbf{\begin{tabular}[c]{@{}c@{}}Description\\ Foreground\\ AVG (G, P, T)\end{tabular}}
        & \textbf{\begin{tabular}[c]{@{}c@{}}Description\\ Background\\ AVG (G, P, T)\end{tabular}}
        & \textbf{\begin{tabular}[c]{@{}c@{}}Transcription\\ Foreground WER\\ $\downarrow$\end{tabular}}
        & \textbf{\begin{tabular}[c]{@{}c@{}}Transcription\\ Background WER\\ $\downarrow$\end{tabular}}
        & \textbf{\begin{tabular}[c]{@{}c@{}}Summarization\\ Foreground\\ ROUGE‑L \,/ METEOR\end{tabular}}
        & \textbf{\begin{tabular}[c]{@{}c@{}}Summarization\\ Background\\ ROUGE‑L \,/ METEOR\end{tabular}}
        & \textbf{\begin{tabular}[c]{@{}c@{}}Free Q\&A\\ Foreground\\ ROUGE‑L \,/ METEOR\end{tabular}}
        & \textbf{\begin{tabular}[c]{@{}c@{}}Free Q\&A\\ Background\\ ROUGE‑L \,/ METEOR\end{tabular}} \\
      \midrule
      Babble     & 87.1   & 86.5   & 10.7  & 29.0  & 59.5 \,/ 56.5 & 44.8 \,/ 43.6 & 65.5 \,/ 68.4 & 63.1 \,/ 63.5 \\
      Pedestrian & 91.1   & 91.1   & 17.6  & 36.7  & 57.5 \,/ 57.2 & 41.2 \,/ 41.8 & 61.1 \,/ 61.4 & 53.6 \,/ 55.4 \\
      \bottomrule
    \end{tabular}%
  }
\end{table*}

\subsection{Impact of Window Size on Correct Speaker Selection}
\label{app:window}

We also examined how varying the window size affects speaker selection accuracy, particularly in dynamic attention scenarios, with our speaker decoding method. The results, detailed in Figure~\ref{fig:window_size}, demonstrate a clear trade-off: shorter windows offer rapid responsiveness but lower accuracy, whereas longer windows significantly improve decoding reliability at the expense of increased latency. These insights suggest the importance of adaptive strategies for window sizing tailored to specific real-time application demands.

\begin{figure}[htbp]
  \centering
  \begin{tikzpicture}
    \begin{axis}[
      width=0.45\textwidth,
      xlabel={Window Size (s)},
      ylabel={Accuracy (\%)},
      symbolic x coords={0.1,0.2,0.5,1,2,4,8},
      xtick=data,
      ymajorgrids=true,
      grid style=dashed,
      mark options={solid},
      ]
      \addplot[
        mark=o,
        thick
      ] coordinates {
        (0.1,34.8)
        (0.2,52.2)
        (0.5,69.2)
        (1,77.6)
        (2,81.4)
        (4,85.6)
        (8,90.2)
      };
    \end{axis}
  \end{tikzpicture}
  \caption{Speaker selection accuracy as a function of window size.}
  \label{fig:window-accuracy}
\end{figure}

\subsection{Ablations}
\label{app:ablations}

To better understand the contributions of different components and design choices in AAD-LLM, we conduct an ablation study with results presented in Table \ref{tab:ablation}. We investigate the impact of the LoRA rank, the necessity of core components, and the type and clustering of features used for attention decoding.å

\noindent
\textbf{LoRA Rank}: We trained the model with smaller ranks of 32 or 128. Increasing the rank from 32 to 512 improves performance in nearly all tasks, particularly for transcription and summarization. 

\noindent
\textbf{Core Components}: Removing Chain-of-thought prompt significantly degrades performance, with WER increasing from 14.4 to 34.9 and ROUGE-L for summarization dropping from 58.3 to 38.3. Removing the separator also degrades performance, increasing WER to 41.8 and reducing summarization ROUGE-L to 40.5. These highlight the crucial role of both components.

\noindent
\textbf{Feature Type and Clustering}: We found that using 8 clusters of x-vectors yielded the best performance on our dataset. Fewer clusters made speaker label prediction easier but often caused the foreground and background speakers to share the same label, making it impossible for the model to distinguish between them. On the other hand, increasing the number of clusters made speaker label prediction more challenging, uniformly degrading downstream task performance. We also tested alternative speaker representations, including speaking style vectors from the StyleTTS 2 \cite{styletts2} text-to-speech model and random vectors from a randomly initialized x-vector extractor, but both performed significantly worse than our chosen x-vectors.

\section{Analysis} \label{app:analysis}

\begin{figure*}[h]
    \centering
    \includegraphics[width=\linewidth]{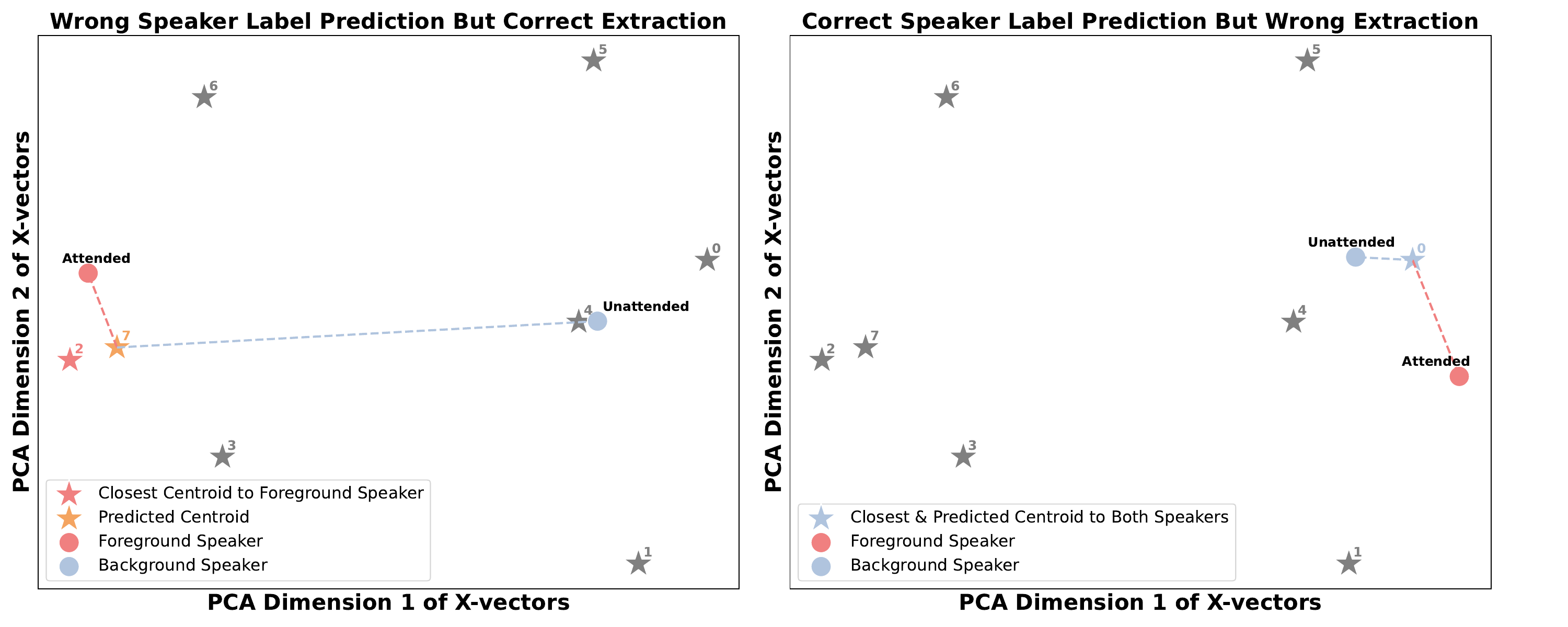}
    \caption{In the left figure, the two speakers have distinct voice characteristics, so even with an incorrectly predicted speaker label, AAD-LLM can still attend to the correct speaker. In contrast, in the right figure, the two speakers have very similar voice characteristics, making it difficult for AAD-LLM to distinguish between the foreground and background speakers—even with a correctly predicted speaker label.}
    \label{fig:analysis}
\end{figure*}

\subsection{Speaker X-vectors and Labels} 
\label{app:xvect}

AAD-LLM utilizes x-vectors to generate speaker identity features for speech clustering. These features enable the creation of comparable speaker profiles and help the model group speakers based on their acoustic similarity. The distribution of x-vectors, visualized using t-SNE \cite{tsne}, is shown in Figure \ref{fig:xvectors}, with the clinical dataset on the left and the speech-only dataset on the right.

In the clinical dataset, there are a total of 280 sentences, with colors representing different speakers. The clusters corresponding to sentences from the eight speakers are clearly distinguishable, demonstrating the effectiveness of x-vectors in speaker separation. In the speech-only dataset, which contains 10,000 sentences, two distinct clusters emerge, corresponding to male and female speakers. The plot is color-coded based on K-means clustering with eight clusters, which are evenly distributed between the two genders.

In addition to x-vectors, Style vectors from StyleTTS2 \cite{styletts2} were also evaluated for their ability to differentiate speakers, as shown in Table \ref{tab:ablation}. However, since x-vectors provided the best performance for intention decoding, they were selected for the final model.

\begin{table}[t]
    \caption{AAD-LLM can attend to the correct speaker in case the speaker label prediction is wrong, but the predicted speaker centroid is still closer to the attended than to the unattended speaker. The results correspond to the speakers in the left part of Figure \ref{fig:analysis}.}
    \centering
    \begin{adjustbox}{width=\columnwidth,center}
    \begin{tabular}{|c|c|c|c|}
        \hline
        Predicted Label & Prediction & Extraction & Transcription \\
        \hline
        0 & \ding{55} & \ding{55} & ``American American Movies.''\\
        \hline
        1 & \ding{55} & \ding{55} & ``American American Movies.''\\
        \hline
        2  & \ding{51} & \ding{51} & ``The trip lasted ten days.''\\
        \hline
        3 & \ding{55} & \ding{51} & ``The trip lasted ten days.''\\
        \hline
        4 & \ding{55} & \ding{55} & ``American American Movies.''\\
        \hline
        5 & \ding{55} & \ding{55} & ``American American Movies.''\\
        \hline
        6 & \ding{55} & \ding{51} & ``The trip lasted ten days.''\\
        \hline
        7 & \ding{55} & \ding{51} & ``The trip lasted ten days.''\\
        \hline
    \end{tabular}
    \end{adjustbox}
    \label{tab:good_analysis}
\end{table}

\subsection{Wrong Predicted Speaker Label but Correct Decoded Speaker} \label{app:mani}

There are two related classification tasks and accuracies for AAD-LLM. One is the speaker label prediction task and the prediction accuracy, which is 78.4\% averaged across five runs in the clinical testing set. The other one is the auditory attention decoding (AAD) task and the AAD / speech extraction accuracy, which is 94.4\% averaged across five runs (Table \ref{tab:aad}).

The difference in accuracy arises from the difference in the tasks. The speaker prediction involves classifying one of K=8 possible speaker labels, an intermediate step, whereas the extraction accuracy reflects the final selection of the attended speaker—determining which speech signal is extracted to the listener’s ears. This selection also impacts the performance of other tasks presented in Table \ref{tab:main}.

The AAD accuracy (94.4\%) is typically higher than the speaker prediction accuracy (78.4\%). This is because we have 8 clusters for two speakers, and even if the speaker prediction is incorrect, the wrong cluster centroid may still be closer to the attended speaker than to the unattended one, eventually leading to the correct speaker selection. A clear example of this is illustrated in the left part of Figure \ref{fig:analysis}, where the model incorrectly predicts the speaker centroid $v_7$ instead of the closest centroid $v_2$. However, $v_2$ remains closer to the foreground speaker than to the background speaker. Additionally, Table \ref{tab:good_analysis} shows that AAD-LLM correctly performs foreground speech extraction and transcription tasks when using four of the eight speaker centroids as inputs, as all of these centroids are positioned closer to the attended speaker than to the unattended one.

\subsection{Failure Cases} \label{app:failure}

AAD-LLM relies on speaker prediction for intention decoding (Section \ref{sec:decoding}), which can become challenging when multiple speakers share similar voice characteristics, such as having the same gender and pitch range. If the speakers are evenly distributed across K speaker clusters, there is a 1/K probability that two distinct speakers will be assigned the same speaker label, making it impossible for AAD-LLM to distinguish between them. As a result, there is a 1/2K chance that AAD-LLM will attend to the incorrect speaker. The right side of Figure \ref{fig:analysis} illustrates this scenario, where both the foreground and background speakers are closest to the same speaker centroid. In this case, the background speaker is even closer, increasing the likelihood that AAD-LLM will mistakenly attend to the wrong speaker.

Increasing the number of clusters can reduce the likelihood that two speakers will be assigned to the same cluster. However, this also makes speaker classification more challenging, as it requires more training data and/or cleaner neural signals to achieve reliable performance.

\section{Subjective Evaluation Details}
\label{app:subj}
A total of 40 participants were recruited through Prolific to evaluate model-generated answers. Participants were divided into four batches of ten. The first batch rated responses in the Summarization task, while batches two through four evaluated responses in the Free Q\&A task. Each participant completed 50 trials.

In each trial, participants listened to an audio stimulus containing three components: the target talker, the non-target talker, and background noise. The target and non-target talkers were spatially separated, with one presented in the left ear and the other in the right ear. Background noise was presented diotically in both ears. Participants were instructed to attend to the target talker, which was indicated by an on-screen arrow. The assignment of the target talker to the left or right ear was randomized across trials. Participants were allowed to replay the audio stimulus multiple times.

\begin{figure*}[t]
    \centering
    \includegraphics[width=\textwidth]{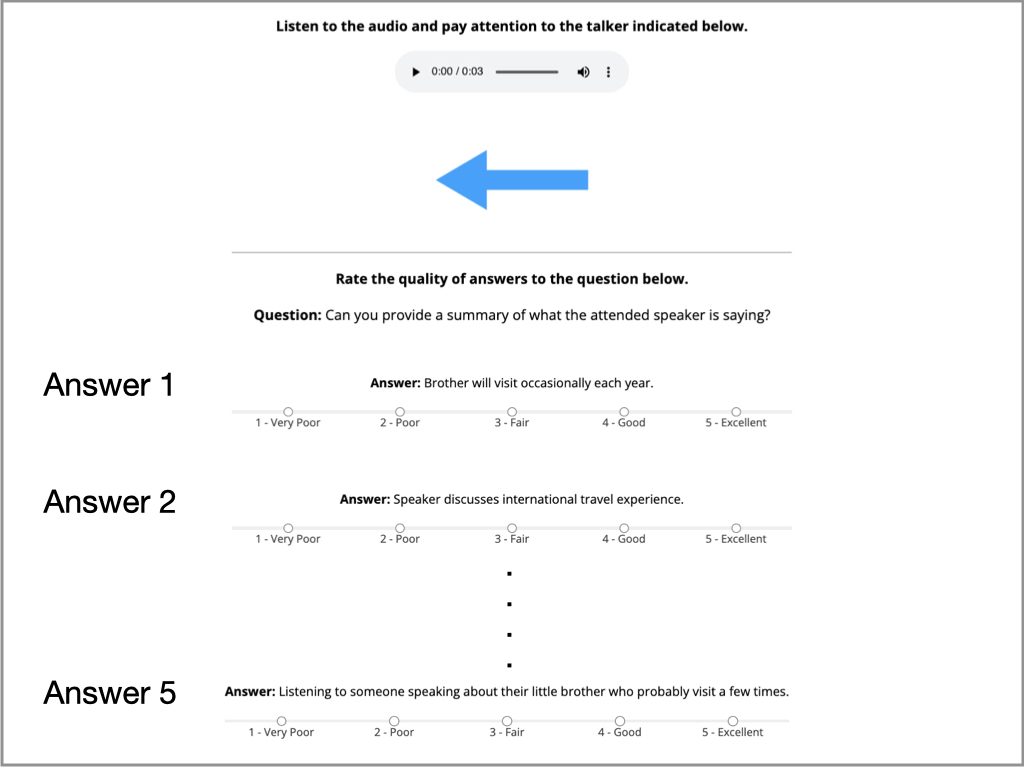}
    \caption{A screenshot from the psychcophysics task where human raters were asked to rate answers from different models (blind and in random order).}
    \label{fig:psychoFig}
\end{figure*}

On the same trial page, participants were presented with a question alongside five model-generated responses, each prompted with the same question. They were asked to rate each response on a 5-point scale, where 1 = "Very Poor" and 5 = "Excellent". Responses containing information from the non-target talker were explicitly instructed to be rated poorly. The presentation order of the model-generated answers was randomized to minimize bias.

All participants were required to wear stereo headphones or earphones and passed an initial left-right channel check to ensure proper perception of spatial separation of the talkers. The total task duration was approximately 30 minutes, and each participant was compensated \$10 for their time. Participants were native English speakers based in the United States, aged between 18 and 40, with no reported hearing difficulties or cognitive impairments. Additionally, all participants had a 100\% approval rating on prior Prolific tasks. A trial screenshot from the psychophysics task is shown in Figure~\ref{fig:psychoFig}.

\begin{table*}[t]
\centering
\caption{Post-hoc pairwise comparisons of human ratings for the Summarization task.}
\renewcommand{\arraystretch}{1.5}
\begin{tabular}{|m{80pt}<{\centering}|m{80pt}<{\centering}|c|c|c|}
\hline
\textbf{Group 1} & \textbf{Group 2} & \textbf{p-value} & \textbf{Bonferroni p-value} & \textbf{Significance} \\
\hline
Oracle Speaker + Qwen2-Audio & Random Speaker + Qwen2-Audio & <0.001 & <0.001 & *** \\
\hline
Oracle Speaker + Qwen2-Audio & Qwen2-Audio & <0.001 & <0.001 & *** \\
\hline
Oracle Speaker + Qwen2-Audio & AAD-LLM w/ brain-decoded attention & <0.001 & 0.002 & ** \\
\hline
Oracle Speaker + Qwen2-Audio & AAD-LLM w/ oracle attention & <0.001 & 0.006 & ** \\
\hline
Random Speaker + Qwen2-Audio & Qwen2-Audio & 0.09 & 0.99 & ns \\
\hline
Random Speaker + Qwen2-Audio & AAD-LLM w/ brain-decoded attention & <0.001 & <0.001 & *** \\
\hline
Random Speaker + Qwen2-Audio & AAD-LLM w/ oracle attention & <0.001 & <0.001 & *** \\
\hline
Qwen2-Audio & AAD-LLM w/ brain-decoded attention  & <0.001 & <0.001 & *** \\
\hline
Qwen2-Audio & AAD-LLM w/ oracle attention & <0.001 & <0.001 & *** \\
\hline
AAD-LLM w/ brain-decoded attention & AAD-LLM w/ oracle attention & 0.79 & 1.00 (capped) & ns \\
\hline
\end{tabular}
\label{tab:stats_comparison_summ}
\end{table*}

\begin{table*}[t]
\centering
\caption{Post-hoc pairwise comparisons of human ratings for the Free Q\&A task.}
\renewcommand{\arraystretch}{1.5}
\begin{tabular}{|m{80pt}<{\centering}|m{80pt}<{\centering}|c|c|c|}
\hline
\textbf{Group 1} & \textbf{Group 2} & \textbf{p-value} & \textbf{Bonferroni p-value} & \textbf{Significance} \\
\hline
Oracle Speaker + Qwen2-Audio & Random Speaker + Qwen2-Audio & <0.001 & <0.001 & *** \\
\hline
Oracle Speaker + Qwen2-Audio & Qwen2-Audio & <0.001 & <0.001 & *** \\
\hline
Oracle Speaker + Qwen2-Audio & AAD-LLM w/ brain-decoded attention & 0.004 & 0.042 & * \\
\hline
Oracle Speaker + Qwen2-Audio & AAD-LLM w/ oracle attention & 0.002 & 0.019 & * \\
\hline
Random Speaker + Qwen2-Audio & Qwen2-Audio & 0.037 & 0.37 & ns \\
\hline
Random Speaker + Qwen2-Audio & AAD-LLM w/ brain-decoded attention & <0.001 & <0.001 & *** \\
\hline
Random Speaker + Qwen2-Audio & AAD-LLM w/ oracle attention & <0.001 & <0.001 & *** \\
\hline
Qwen2-Audio & AAD-LLM w/ brain-decoded attention & <0.001 & <0.001 & *** \\
\hline
Qwen2-Audio & AAD-LLM w/ oracle attention &<0.001 & <0.001 & *** \\
\hline
AAD-LLM w/ brain-decoded attention & AAD-LLM w/ oracle attention & 0.80 & 1.00 (capped) & ns \\
\hline
\end{tabular}
\label{tab:stats_comparison_free}
\end{table*}

\end{document}